\documentclass{aa}  
\usepackage{xcolor}
\usepackage{graphicx}
\usepackage{txfonts}
\usepackage{lipsum}
\usepackage{subcaption}         
\usepackage{lscape}             
\usepackage{placeins}           
                                
\usepackage{soul} 

\begin{document}

   \title{Extragalactic microlensing through Ultra Diffuse Galaxies}


%
%
%

   \author{Sung Kei Li\inst{1, 2, 3}\corrauth{skli@ifca.es}\fnmsep\thanks{Croucher fellow}        
        \and Thomas Broadhurst\inst{4, 5, 6}\email{tom.j.broadhurst@gmail.com}
        \and Jose M. Diego\inst{1}\email{jdiego@ifca.unican.es}
        \and Jeremy Lim \inst{2, 3}\email{jjlim@hku.hk}
        \and Jose M. Palencia \inst{1}\email{palencia@ifca.unican.es}
        \and James Nianias \inst{2, 3}\email{nianias@connect.hku.hk}
        \and Jiashuo Zhang \inst{5}\email{jiashuo.zhang@dipc.org}
        }

   \institute{IFCA, Instituto de F\'isica de Cantabria (UC-CSIC), Av. de Los Castros s/n, 39005 Santander, Spain  \and Department of Physics, The University of Hong Kong, Pokfulam Road, Hong Kong  \and The Hong Kong Institute for Astronomy and Astrophysics, The University of Hong Kong, Pokfulam Road, Hong Kong, P. R. China \and Department of Theoretical Physics, University of Basque Country UPV/EHU, Bilbao, Spain \and Donostia International Physics Center, Paseo Manuel de Lardizabal, 4, San Sebasti\'an, 20018, Spain \and Ikerbasque, Basque Foundation for Science, Bilbao, Spain}

   \date{\today}

 
  \abstract{Stellar microlensing is a powerful method to constrain compact dark matter models, uncover binary stars, and exoplanets during caustic crossing events. At cosmological distances, {\it James-Webb Space Telescope} ({\it JWST}) is routinely detecting microlensed giant stars in highly magnified galaxies behind massive lensing clusters. Here, we explore for the first time microlensing in modest redshift galaxies commonly seen through local Ultra Diffuse Galaxies (UDGs). Using the UDG NGC1052-DF2 as a proof-of-concept, we find that detecting microlensing events through UDGs is possible. Euclid is ideal for identifying samples of low-redshift star-forming galaxies seen through local galaxies for deeper cadenced follow-up, where our zeroth-order calculation estimates that $\mathcal{O}(1-10)$ events per year are expected over the whole sky under the monitoring of LSST, assuming all UDGs share similar distance and stellar content as NGC1052-DF2. Such measurements can help settle the DM issue for UDG galaxies by providing a stellar mass estimate weighted more by the low end of the stellar initial mass function (IMF) than standard stellar synthesis estimates. Note, lower efficiency is estimated for {\it JWST} of $\sim 5.6\times10^{-2}\,\textrm{yr}^{-1}$ over its five background galaxies is expected for typical {\it JWST}$\sim29$\,mag visits, and a low Vera Rubin Legacy Survey of Space and Time (LSST) detection rate of$\sim 0.03\,\textrm{yr}^{-1}$ such that NGC1052-DF2 might not be a prime target given its lack of low-redshift background galaxies.}    


   \maketitle

\nolinenumbers

\section{Introduction}

Galactic microlensing, first detailed by \citet{Liebes_1964, Paczynski_1986}, has been used to constrain the abundance of massive compact halo objects in dark matter haloes \citep[e.g.,][]{Wyrzykowski_2011, Calcino_2018, Niikura_2019_PBH, Montero_Camacho_2019, Mroz_2024}, probe galactic structures and stellar properties \citep[e.g.,][]{Novati_2008, Moniez_2010, Mroz_2019, Rodriguez_2022}, and discover exoplanets \citep[e.g.,][]{Abe_2004, Gaudi_2012, Tsapras_2018, Mroz_2024_exo}. While collaborations such as EROS \citep[e.g.,][]{Afonso_1999} and OGLEs \citep[e.g.,][and thereafter]{Udalski_1994} have detected tens of thousands of events in our Milky Way, Small and Large Magellanic Clouds over the past few decades, confirmed detections are so far available only as far as the M31 galaxy \citep[through the Subaru telescope; see, e.g.,][]{Niikura_2019}, whereas candidate events go as far as the M87 galaxy \citep{Baltz_2004}. At cosmological distances, individual stars can only be detected when microlensing is augmented by strong lensing by foreground galaxy clusters \citep[e.g.,][]{Kelly_2018_Icarus, Kelly_2022_Flashlights, Yan_2023, Fudamoto_2025, palencia2026statisticalstudy100magnified}, and speculatively, perhaps some subsamples of galaxy-galaxy strong lensing systems \citep[e.g.,][]{Li_2025_Horseshoe}. The study of the spatial distribution and detection rate of lensed stars provides unique insights into, for example, the nature of dark matter \citep[][]{Williams_2024, Diego_2024_3M, Broadhurst_2025} and star formation physics beyond the local universe \citep[e.g.,][]{Diego_2024_BF, Li_2025_IMF}. 

In this paper, we propose to make use of UDGs as a semi-transparent, thin lens plane that supplies microlenses to detect individual distant stars in background galaxies. The concept of UDGs was first introduced in \citet{Sandage_1984} but only formally named much later in \citet{Van_Dokkum_2015} -- observations reveal galaxies with extremely low surface brightness \citep[$\gtrsim 24\,$mag$\,\textrm{arcsec}^{-2}$,][]{Chamba_2020}, that have Milky Way-like size but mass comparable to dwarfs \citep{Benavides_2021}. Recent studies suggest that some UDGs may lack dark matter content entirely \citep{van_Dokkum_2018, keim2026galaxymissingdarkmatter}, making their formation channel particularly intriguing \citep[e.g., through high-speed collision,][]{Silk_2019, Lee_2024} or hinting towards alternative dark matter models \citep[e.g.,][]{Yang_2020, Pozo_2021}

\begin{figure}[t!]
    \centering
    \includegraphics[width=\linewidth]{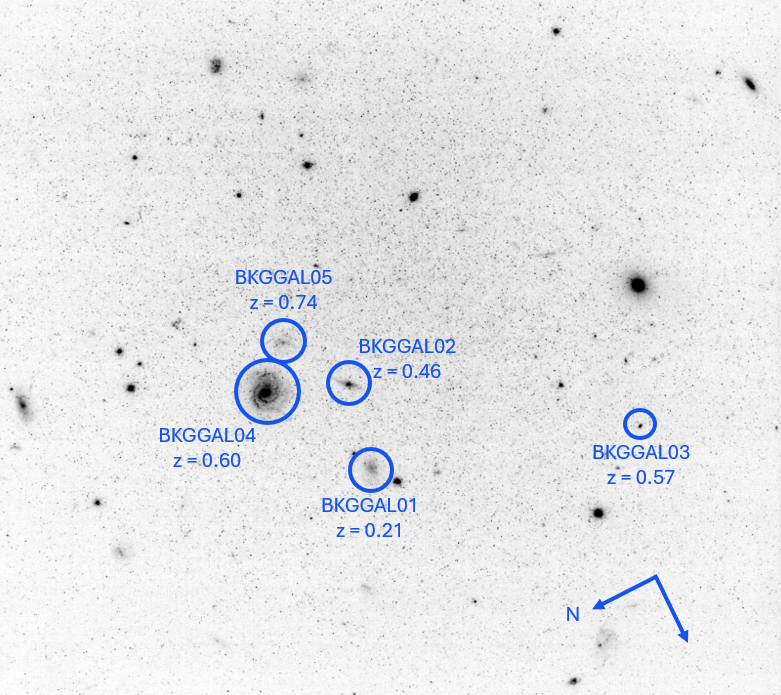}
    \caption{Negative {\it JWST} F090W image of NGC1052-DF2 (GO-3990, PI: Morishita). Five spectroscopically confirmed background galaxies are labelled with their redshifts. We carry out event rate analysis on these galaxies based on their H-$\beta$ luminosities, as one shall read later in Section~\ref{sec: rate}. }
    \label{fig: NGC1052-DF2}
\end{figure}

In the context of microlensing, the low surface brightness (and stellar mass density) of UDGs facilitates the detection of background galaxies without being obscured by foreground light. 
Their low mass and diffuse nature also do not create any known strong gravitational lensing effect (surface mass density much lower than the critical surface mass density for all source redshifts), such that only the microlensing effect has to be considered and the macroscopic model can be safely ignored. We make use of the UDG NGC1052-DF2, as shown in Fig.~\ref{fig: NGC1052-DF2}, as a proof-of-concept in this paper, where we first consider the ideal case and see if UDG microlensing events are theoretically detectable in Section~\ref{sec: max_mu}. We then estimate the relevant time scales of UDG microlensing in Section~\ref{sec: timescale}, and the expected event rate for different possible scenarios in Section~\ref{sec: rate}. We finally discuss caveats and possible applications of UDG microlensing in Section~\ref{sec: discuss}. Throughout this work, we adopt the AB magnitude system \citep{Oke_1983}, along with Planck18 cosmological parameters: $\Omega_{m} = 0.31$, $\Omega_{\Lambda} = 0.69$, and $H_{0} = 67.4\,\textrm{km s}^{-1}\, \textrm{Mpc}^{-1}$\citep{Planck18}.

\section{Maximum Magnification}

\label{sec: max_mu}

To investigate whether detecting background stars at distant galaxies through UDG microlensing is theoretically possible, we can consider the maximum magnification any given star can attain at corresponding distances. The Einstein radius at image plane, $R_{E}$, given a point-like microlens with mass, $M$, is:

\begin{equation}
    R_{E} = \sqrt{\frac{4GM}{c^{2}}\frac{D_{L}D_{LS}}{D_{S}}},
\end{equation}

\noindent with $D_{S}$ the angular diameter distance to the source, $D_{L}$ the angular diameter distance to the lens, and $D_{LS}$ the angular diameter distance between the lens and the source. $G$ is the gravitational constant, and $c$ the speed of light. For a perfect alignment between a source with radius $r_{s}$ and a single microlens, the maximum magnification, $\mu_{max}$ can be approximated as:

\begin{equation}
    \mu_{max} = A_{i}/A_{s} \approx \frac{2\pi R_{E}\cdot r_{s}}{\pi r_{s}^{2}} = 2\frac{R_{E}}{r_{s}},
\label{eqn: mu_max}
\end{equation}

\noindent for the limit of $r_{s}<< R_{E}$, where $A_{i}$ is the area of the Einstein ring image and $A_{s}$ the area of the source before lensing. For our case, this is explicitly true as $R_{E}$ is much larger than the maximum source radii we shall consider, $r_{s} = 1000\,R_{\odot}$, for all source distances considered. Also, UDG NGC1052-DF2 has a mean surface mass density, $\Sigma_{\star}$ of $\sim 5\,M_{\odot}/\textrm{pc}^{2}$ according to multiple analyses \citep{van_Dokkum_2018, Trujillo_2019, Shen_2021}. For a typical stellar IMF deduced from local observations, we can assume a mean stellar mass of $\hat{M} = 0.3\,M_{\odot}$ \citep{Kroupa_2003, Chabrier_2003} for a population that has age of $\sim9\,$Gyr \citep{Fensch_2019}, then we would have $\sim 16$ microlenses given $1\,\textrm{pc}^{2}$ of image plane area on the sky. Each of these microlenses has the same Einstein radius, $R_{E}$, which yields the image plane area of $\pi R_{E}^{2}$. We calculate for all source distances, $D_{S}$, considered in this paper such that $\pi R_{E}^{2}  \color{black} \times \Sigma_{\star}/\hat{M} << 1\,$. This means that the image plane area is far from fully covered by microlensing Einstein rings, such that the single-lens approximation is reasonable in considering at least the ideal case of maximum magnification attainable. We shall review the validity and applications of this approximation later on in Section~\ref{sec: caustic}.

The inferred distance to UDG NGC1052-DF2 is $\sim 20\,$Mpc \citep{van_Dokkum_2018}\footnote{Recent claims from \citet{Trujillo_2019, Beasley_2025, tang2026newmeasurementsdistancesgalaxies} inferred a shorter distance of $\sim 13-17\,$Mpc, which do not significantly affect our inference as the difference in distance is minor for our lensing calculation}. We retrieved the Very Large Telescope Multi-Unit Spectroscopic Explorer (MUSE) data cubes associated with ESO-DDT programs 2101.B-5008(A) and 2101.B-5053(A) (PI: Emsellem) and reduced the spectroscopic redshifts of the background sources behind NGC1052-DF2 based on their H$\beta$ emissions, as shown in Fig.~\ref{fig: NGC1052-DF2}. To be more informative, we hereby also consider a suite of source distances in our analysis for extrapolating the case beyond NGC1052-DF2. For a comprehensive analysis, we consider three characteristic masses of stellar microlenses: $0.01\,M_{\odot}$, $0.1\,M_{\odot}$, and $1\,M_{\odot}$, and a few source radii: $1\,R_{\odot}$, $10\,R_{\odot}$, $100\,R_{\odot}$, and $1,000\,R_{\odot}$. The Fresnel scale, $R_{F} = \sqrt{\lambda \, \frac{D_{L}D_{LS}}{D_{S}}}$, is $\sim1.1\,R_{\odot}$ for lens at 22 Mpc with source as far as 1300 Mpc ($z \approx 0.5$), and $0.8\,R_{\odot}$ for source as close as 44\,Mpc, assuming wavelength, $\lambda$, of 1 micron. Since $R_{E}>>R_{F}$ for all cases, even considering the effect of magnification (which unit length scales as $\sqrt{\mu}$), UDG microlensing events are well within the geometric lens regime without having to consider wave optics.

\begin{figure}
    \centering
    \includegraphics[width=\linewidth]{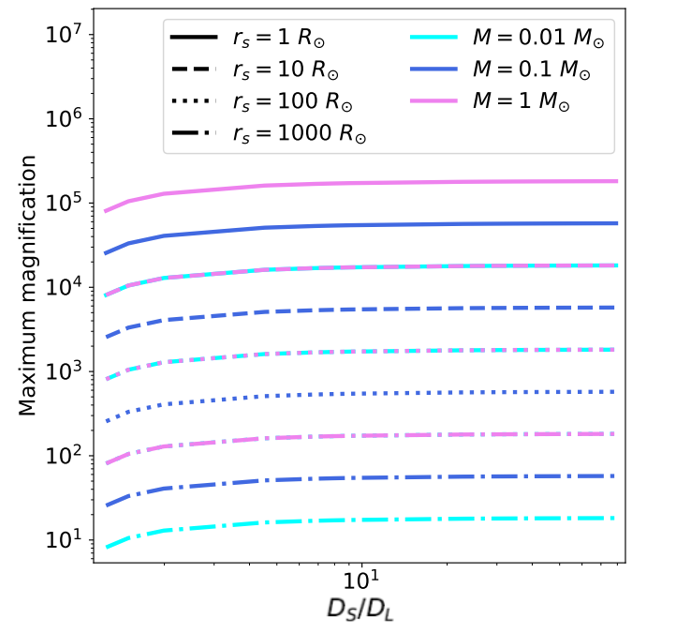}
    \caption{Maximum magnification attainable considering the case of NGC1052-DF2 against different source angular diameter distances in terms of $D_{S}/D_{L}$, given different source radii represented by the line styles, and different masses of microlenses represented by the colors.}
    \label{fig: max_mu}
\end{figure}

We show the maximum magnification attainable for all combinations of source distances, mass of stellar microlenses, as well as the source radii in Fig.~\ref{fig: max_mu}, calculated through Eqn~\ref{eqn: mu_max}. One can see that for a small source with $\sim 1 R_{\odot}$, it can attain extreme magnification up to $\sim 10^{5}$ with a massive microlens of $\sim 1 M_{\odot}$ under perfect alignment. The condition of brightness of stars against the stellar radii is constrained by stellar physics. As a simple calculation, we adopt the MESA Isochrone and Stellar Tracks \citep[MIST,][]{Choi_2016} to see what the brightest star one can see is, given the considered range of source radii. For the analysis hereafter, the detection limit we referred to is always the rest-frame i-band (roughly corresponds to the {\it JWST} F090W) magnitudes compiled by MIST, unless specified, to avoid doing the K-correction every time.

\begin{figure}
    \centering
    \includegraphics[width=\linewidth]{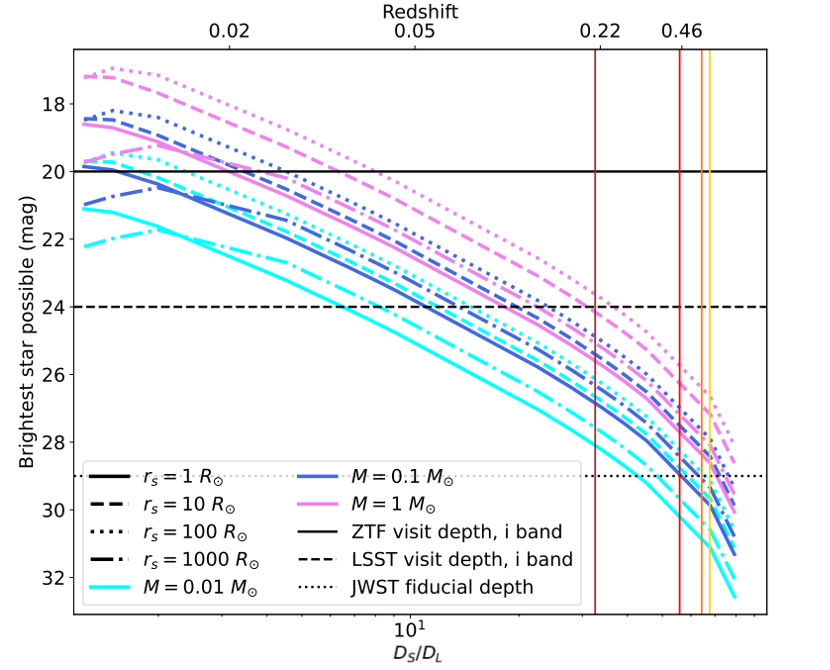}
    \caption{Brightest star possible (in apparent magnitudes) under ideal situations at different source angular diameter distances and redshifts (given that our $D_{L}$ is sufficiently small, $D_{L} - D_{S}$ is sufficiently close to $D_{LS}$ down to $\lesssim1\% $level), considering the maximum magnification shown earlier in Fig.~\ref{fig: max_mu}. Since luminous stars cannot have arbitrarily small radii, we consider the brightest stars that have each of the stellar radii allowed by the MIST isochrones. The two horizontal lines represent the visit depths of ZTF and LSST, respectively. The vertical lines denote the redshifts of the sources behind NGC1052-DF2, as shown earlier in Fig.~\ref{fig: NGC1052-DF2}. }
    \label{fig: brightest_star}
\end{figure}

From Fig.~\ref{fig: brightest_star}, one can see that background stars undergoing UDG microlensing can get as bright as $\sim 18\,$mag for a nearby source. We have also placed the redshift of the few background galaxies behind NGC1052-DF2 in vertical lines with different colors as a reference. The depths of ZTF ($\sim 20\,$mag), LSST ($\sim 24\,$mag) all-sky survey, and a fiducial 2-hour exposure {\it JWST} depth of $\sim 29\,$mag are denoted as black lines with different linestyles. Notice that this is a single point source detection threshold. For transient signals picked up from image differencing, the photon noise from the background galaxies would have contributed to the uncertainty, such that the detection limit would have been dimmer depending on the brightness of the source galaxies. Our calculation hereafter is thus an upper limit of the event rate. This simply implies that, for an ideal scenario, detecting lensed stars at distant galaxies is statistically possible. For the case of NGC1052-DF2, LSST could, in theory, capture events in one of the galaxies (BKGGAL01, $z = 0.2$). Using {\it JWST} will allow one to observe events for background galaxies even further away.

\section{Relevant Timescales}
\label{sec: timescale}




Unlike galactic microlensing events, where the background stars are always detectable, background stars with apparent magnitude $m$ in UDG microlensing are only detectable when they have attained some minimum magnification:

\begin{equation}
    \mu_{min}(m) \geq 10^{(m-m_{5\sigma})/2.5},
\end{equation}

\noindent under a detection limit of $m_{5\sigma}$. The light curve scale is then the duration when the stars have attained sufficient magnification to be detectable. The magnification of the background source with radii much smaller than the Einstein radius at distance $r$ from the caustic center of microlens is given as:

\begin{equation}
    \mu = \frac{b^{2}+ 2}{b\sqrt{b^{2}+4}},
\label{eqn: mag_vs_b}
\end{equation}

\noindent with $b$ the impact parameter defined on source plane given as:

\begin{equation}
    b = \frac{r}{ \tilde{R_{E}}} = \frac{r}{ R_{E}D_{S}/D_{L}} ,
\label{eqn: b}
\end{equation}

\noindent for the limit where $r>> r_{s}$. We can reverse Eqn~\ref{eqn: mag_vs_b} to find $b$ where the star begins to be detectable (at $\mu_{min}$):

\begin{equation}
    b(\mu) = \sqrt{\frac{2\mu}{\sqrt{\mu^{2}-1}}-2} \approx \sqrt{\frac{1}{\mu^2} + \cdots}\approx \frac{1}{\mu} \;\; \text{for } \mu \gg 1.
\end{equation}

For perfect alignment such that the star enters the Einstein radius in the radial direction, the total time taken would be: 

\begin{equation}
    t_{b} \approx 2\times \frac{|r_{s} - b(\mu_{min})\times \tilde{R_{E}}|}{v}
\label{eqn: t_b}
\end{equation}

\noindent where the maximum magnification is topped at the limit where $r \approx r_{s}$ as we have derived earlier in Eqn~\ref{eqn: mu_max}. $v$ is the relative transverse velocity, which is dominated by the peculiar motion owing to the Hubble flow. With NGC1052 DF2's distance of $\sim 20\,$Mpc, this would be approximately $\sim 400\,\textrm{km}/\textrm{s}$ for a distant source \citep[e.g.,][]{Hudson_2004, Ma_2011}. 
For a lensed star with a source radius of $1000R_{\odot}$ that has $-9\,$mag absolute magnitude as constrained by the isochrone, $\mu_{min}$ would be $\sim 100$ considering redshift $\sim 0.6$ and observation depth of $\sim 29\,$mag. For a $0.3M_{\odot}$ lens, $R_{E} = 5\times10^{4}R_{\odot}$. Solving for $t_{b}$ gives $\sim 20\,$days, which means the light curve of such a scenario can be captured with LSST's 3-day cadence, or by dedicated {\it JWST} observations that have even higher temporal resolution.






\section{Estimated event rate}
\label{sec: rate}

The previous analysis in Section~\ref{sec: max_mu} is ideal, demanding the perfect alignment of stars, and just to prove that it is possible to detect extragalactic microlensing through UDGs. Here, we estimate the number of events expected for UDG microlensing.

The classic \citet{Paczynski_1986} microlensing optical depth given some lensing stellar population with three-dimensional mass density, $\rho$, is:

\begin{equation}
    \tau = \frac{4\pi G}{c^{2}D_{S}} \int^{D_{S}}_{0}\rho(D_{L})D_{L}D_{LS}\, dD_{L}.
\end{equation}

\noindent Since the UDG is sufficiently thin given $D_{S}$, $D_{L}$, and $D_{LS}$, and stars from the UDG are the primary source of microlenses, the condition can be simplified without integrating along the line-of-sight:

\begin{equation}
    \tau = \frac{4\pi G}{c^{2}}\Sigma_{\star} \frac{D_{L}D_{LS}}{D_{S}},
\label{eqn: tau_2}
\end{equation}

\noindent with $\Sigma_{\star}$ the surface mass density of stellar microlenses in the UDG.

The optical depth, $\tau$, represents the probability that any random source is inside the Einstein radius of a microlens at any given time when the source size is smaller than the Einstein radius. Considering $N$ stars in the background galaxies, then it simply means that we can expect $N\times\tau$ stars that are inside the Einstein radius of some microlenses if the whole background galaxy is being covered by the thin sheet of microlenses. Now, for this case, the stars will be unlikely to attain the maximum magnification introduced in Eqn~\ref{eqn: mu_max} since that demands a perfect alignment. Rather, the magnification they are attaining is related to where exactly they are locating inside the Einstein radius through Eqn~\ref{eqn: mag_vs_b}. For a random pair of source and microlens, the probability of having any impact parameter $P(b)$ should scale with source plane area covered as $\propto  r^{2} \propto b^2$. This would hold until the limit of $r \approx r_{s}$, where the size of the star becomes relevant, and such that maximum magnification is given by Eqn~\ref{eqn: mu_max}. For sources at redshifts $\sim 0.5$ with stellar radii $\sim 1000 R_{\odot}$, this would require an impact parameter of $\sim 10^{-4}$. We can truncate the maximum magnification attainable with Eqn~\ref{eqn: mu_max} at such characteristic impact parameters. The probability density of a star getting some magnification would then be:

\begin{equation}
    \frac{dP}{d\mu} = \frac{dP}{db} \times \frac{db}{d\mu}\propto \mu^{-3}.
\label{eqn: dPdmu}
\end{equation}

To know the expected detection rate, we would need to determine the number of stars in the background source, and their probability of falling into the Einstein radius of a microlens, combined with the probability that they will attain some magnification that is sufficient for them to be detectable with some detection threshold. The first unknown will be the luminosity function of the background galaxy, $N$, where the second probability is described by the optical depth, $\tau$, and the third probability density scales with $\mu^{-3}$ as just introduced.

Since $\tau$ already accounts for the probability that a source falls into $R_{E}$, $b \leq 1$. The probability that the background will attain such magnification is simply $\mu_{min}^{-2}$ upon definite integration of Eqn~\ref{eqn: dPdmu}. Of course, it would not make sense to consider stars that require $\mu_{min}>\mu_{max}$ since this means they would never be detectable as they could never have attained sufficient magnification, limited by their intrinsic source radii. Here, we again assume the mean mass of stellar microlenses to be $0.3\,M_{\odot}$ and compute $\mu_{max}$ for all stars sampled in the luminosity function. The expected number of stars, $E$, one would observe given some detection limit is therefore:

\begin{equation}
    E = \int N(m) \cdot \tau \cdot \int^{\mu_{max}}_{\mu_{min}(m)} \frac{dP}{d\mu} \color{black} d\mu \,dm
\label{eqn: expected}
\end{equation}



To generate the luminosity function, we assume the background galaxies have a constant star formation history, which is the underlying assumption of the \citet{Kennicutt_1983} relation, such that the star formation rate (SFR) can be inferred via the H-$\alpha$ luminosity of background galaxies independently. Although the original \citet{Kennicutt_1983} relation is limited to the last $\sim 10\,$Myrs, we naively extrapolate it to $\sim 100\,$Myrs as the lensed star detection rate is expected to be dominated by star formation episodes over the last $\sim 100\,$Myrs \citep{Li_2025_BRratio}.
This approximation should thus give us a reasonable order-of-magnitude estimation of the event rate expected.
We sample through a \citet{Kroupa_2003} IMF and retrieve the luminosity function (also the corresponding stellar radii for each individual star) following MIST isochrones. This way, we can conveniently estimate how many lensed stars can be detected through this method, as a function of the assumed-constant SFR.

Furthermore, if Eqn.~\ref{eqn: t_b} gives the mean time scale of UDG microlensing light curves, then the expected event rate per year (i.e., the rate by constantly monitoring the source over a year) can be estimated by combining this with the expected event rate in Eqn~\ref{eqn: expected}, given as:

\begin{equation}
    \Gamma (yr^{-1}) = \int \frac{N(m)}{t_{b}} \cdot \tau \cdot \int^{\mu_{max}}_{\mu_{min}(m)} \frac{dP}{d\mu} \color{black} d\mu \,dm,
\label{eqn: peryear}
\end{equation}

\noindent where $t_{b}$ can be computed for each star in the luminosity function, $N(m)$, simulated earlier.

\begin{figure*}
    \centering
    \includegraphics[width=1.\linewidth]{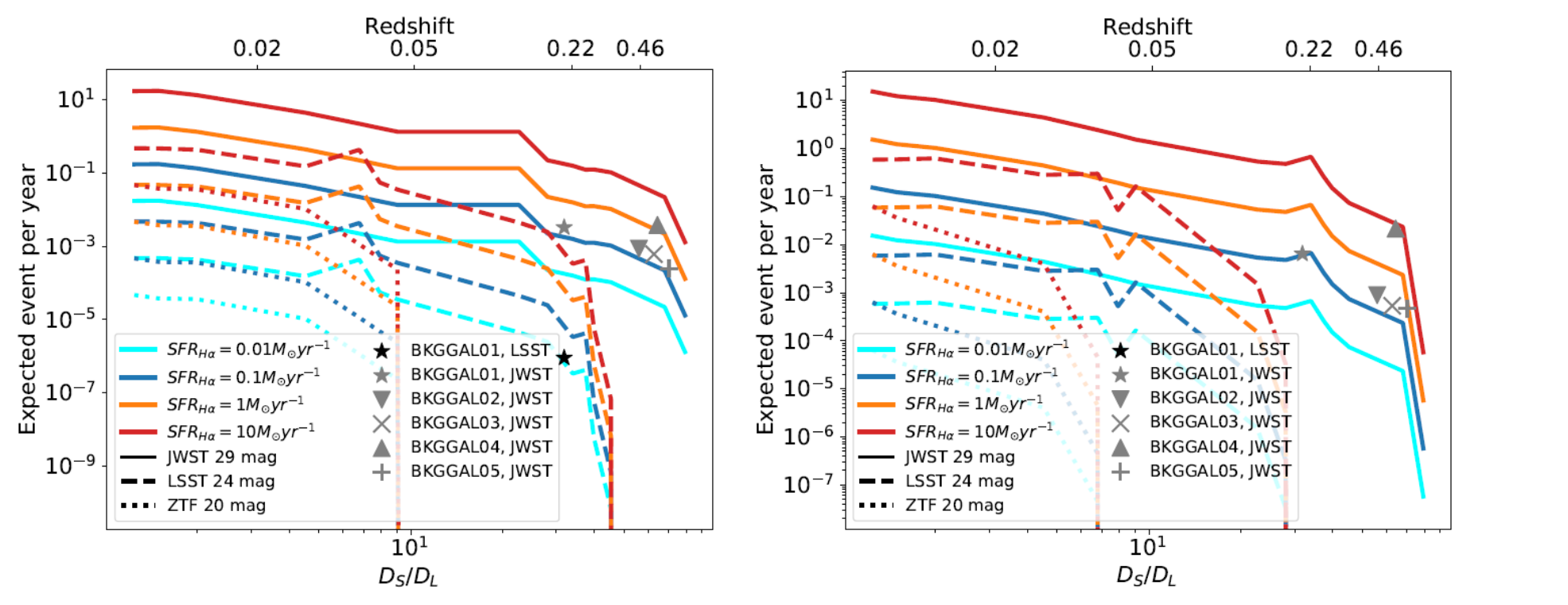}
    \caption{Expected event rate per year, considering NGC1052-DF2 acting as the lens with different sources (with different SFR rate indicated by line colors, assuming a constant star formation history over the last $\sim 100\,$Myrs), at the background at different source angular diameter distances (redshifts). Solid lines refer to calculations made assuming {\it JWST} depth of $\sim 29\,$mag, where dashed and dotted lines represent the calculations assuming LSST and ZTF depths of $\sim 24\,$mag, and $\sim20\,$mag, respectively. Individual data points are the predictions made for each of the background galaxies behind NGC1052-DF2.}
    \label{fig: detect}
\end{figure*}

Here we show the expected event rate per year in Fig.~\ref{fig: detect} assuming a {\it JWST} depth of $29\,$mag (solid lines), LSST depth of $\sim 24\,$mag (dashed lines), and ZTF depth of $\sim 20\,$mag (dotted lines). On the left panel, we show the case for the rest-frame u band, which best captures the UV light dominated by young, massive stars that are usually smaller in radius ($\sim 10\,R_{\odot}$) and able to attain higher maximum magnification; on the right panel, we show the case for the rest-frame i band, that have the highest sensitivity towards red, evolved stars which have higher luminosity but larger radius ($\sim 100\,R_{\odot}$), thus lower maximum magnification. Given that the two bands are far separated in wavelength, it is very likely that stars detected in one band will not be detected in the other, except for stars with moderate surface temperature, like yellow supergiants.
For both cases, we show source galaxies with different SFR in different colours, ranging between $0.01\,M_{\odot}\,\textrm{yr}^{-1}$ and $10M_{\odot}\,\textrm{yr}^{-1}$. For other galaxies, the event rate scales linearly with the SFR since it normalizes the luminosity function to the first order.



One can see that the furthest UDG microlensing event detectable depends a lot on the detection threshold, as it determines whether $\mu_{min} > \mu_{max}$. For ZTF-like observation, the furthest event detectable would be at $\sim 9$ times the lens distance, at a rather local redshift of $\sim 0.05$ for the case of NGC1052-DF2 acting as the lens. For LSST-like observation, this extends to redshift $\sim 0.2$. These limits are also given by the brightest star detectable, as shown earlier in Fig.~\ref{fig: brightest_star}. For the i band (right panel), the furthest detectable events are closer than the case for the u band (left panel), because stars that are IR-bright tend to be evolved stars that have larger radii, limiting the maximum magnification attainable following Eqn~\ref{eqn: mu_max}.
Notice that these ``cliffs'' of non-detection, as shown in Fig.~\ref{fig: detect}, are not scaling linearly with the x-axis (source angular diameter distances) across different instruments (different detection limits) since the detection depends on the luminosity distance, where we convert from the angular diameter distances with its equivalent redshift based on the cosmology adopted. Likewise, the event rate does not directly scale with the depth of the telescopes as the relevant timescale is longer for telescopes with a deeper detection limit (a lower $\mu_{min}$, thus longer $t_{b}$ as shown in Eqn~\ref{eqn: t_b}).

For LSST, one UDG microlensing event in either the u band or the i band is expected for every $\sim 100$ source galaxies with SFR of $\sim 1\,M_{\odot}\textrm{yr}^{-1}$ (or, per $\sim 10^{2} \,M_{\odot}\textrm{yr}^{-1}$ total SFR) at redshift $\sim 0.05$ during the course of a year if the foreground UDG is identical to NGC1052-DF2. The rate would be even higher (or fewer background galaxies to achieve an equivalent detection efficiency) for closer, or even more massive/active star-forming background galaxies, as well as UDGs that are higher in surface mass density.

Here, we carry out an approximate estimate of the expected event rates for each of the low-redshift background galaxies visible through NGC1052-DF2. We do not directly use the H-$\alpha$ luminosity that is conventionally adopted \citep[e.g., in][]{Kennicutt_1998, Kennicutt_2012} as the H-$\alpha$ lines of these background galaxies are either out of the range of MUSE, or fall onto telluric absorptions. Instead, we follow \citet{Zeimann_2014} and convert their H-$\beta$ luminosity to SFR, assuming an intrinsic H$\alpha$/H$\beta$ ratio of $\sim 2.9$ \citep{Brocklehurst_1971}. Limited to the data availability, we do not account for dust extinction, such that the adopted SFR and hence UDG microlensing rate would be an upper limit. We show the SFRs, as well as the expected LSST and {\it JWST} event rate of the five background galaxies as shown in Fig.~\ref{fig: NGC1052-DF2} in Table~\ref{tab: table}. The corresponding rates are also denoted in Fig.~\ref{fig: detect} for reference. We do not repeat the Monte Carlo sampling multiple times to explore the uncertainty in the expected detection rate, given the extremely high computational cost, and that the rate is way lower than one event per year.

Right now, the non-detection of UDG microlensing events in NGC1052-DF2 with existing ZTF data is consistent with the expectation of our methodology, since the background galaxies have redshifts $\gtrsim 0.2$ and detection is forbidden for ZTF as shown in Fig.~\ref{fig: detect}. From Fig.~\ref{fig: detect} and Table~\ref{tab: table}, one can tell that NGC1052-DF2 is probably not a prime target for UDG microlensing detection, as it lacks star-forming galaxies at lower redshifts. The LSST event rate is roughly estimated as $\sim 10^{-6}$ per year. Even with {\it JWST}, the total event rate summed over all five background galaxies is $\sim 0.03$ per year, meaning that only one event is expected over $\sim 30\,$yrs of constant monitoring.

\begin{table*}[]
    \centering
    \begin{tabular}{c|c|c|c|c}
    Background     & z & H$\beta$ SFR & u-band $\Gamma (\textrm{yr}^{-1})$ & i-band $\Gamma (\textrm{yr}^{-1})$\\ 
    galaxy& & $(M_{\odot}\,\textrm{yr}^{-1})$ &  {\it JWST} (LSST) & {\it JWST} (LSST)\\ 
    
    \hline
    BKGGAL01     & 0.21 & $0.06 \pm 0.01\,$ & $\sim3\times 10^{-3}$ ($\sim3\times 10^{-6}$) & $\sim 7\times10^{-3}$ (0) \\
    BKGGAL02     & 0.46 & $0.29 \pm 0.01\,$ & $\sim1\times 10^{-3}$ (0) & $\sim 9\times10^{-4}$ (0)\\
    BKGGAL03     & 0.57 & $0.23 \pm 0.03\,$ & $\sim6\times 10^{-4}$ (0) & $\sim 4\times10^{-4}$ (0)\\
    BKGGAL04     & 0.60 & $1.73 \pm 0.28\,$ & $\sim8\times 10^{-3}$ (0) & $\sim 0.01$ (0) \\
    BKGGAL05     & 0.74 & $0.30 \pm 0.03\,$ & $\sim3\times 10^{-4}$ (0) & $\sim 2\times 10^{-4}$ (0)\\
    \end{tabular}
    \caption{Redshifts, SFR, and inferred UDG microlensing event rate per year in rest-frame u band (left, better capture young blue stars) and rest-frame i band (right, better capture red, evolved stars) for the background galaxies behind NGC1052-DF2.}
    \label{tab: table}
\end{table*}


The latest known catalog of UDGs comes from the DR9 legacy survey, yielding a total of $\sim 7000$ UDG candidates \citep{Zaritksy_2023}. A very quick zeroth-order calculation assuming all these UDGs are identical to NGC1052-DF2 in terms of stellar surface mass density will cover an angular area of $7000 \times \pi \times (8'')^{2}\approx 0.1 \deg^{2}$ \citep[the mean size of UDGs in][]{Zaritksy_2023}. The star formation rate density at the low redshift range of $\lesssim 0.2$, where LSST can detect UDG microlensing events, is $\sim 0.02\, M_{\odot}\textrm{yr}^{-1}c\textrm{Mpc}^{-3}$ \citep{Haslbauer_2023}. Hence, the total star formation rate behind the UDGs will be $\sim\,0.02 M_{\odot}\textrm{yr}^{-1}c\textrm{Mpc}^{-3}\times 0.1 \deg^{2}\times\,6\times10^{4}\,c\textrm{Mpc}^{3}\deg^{-2}$ where the last term is the comoving volume per solid angle between the redshift range of interest. From this approximation, one can expect $\sim 120\,M_{\odot}\textrm{yr}^{-1}$ of total star formation rate behind these galaxies. Under our prediction, the mean detection rate of UDG microlensing at this redshift range would be of the order of $\sim 10^{-2}-10^{-1} \,\textrm{yr}^{-1}$ per $1\, M_{\odot}\textrm{yr}^{-1}$ for LSST u band and i band, yielding a total event rate of $\sim \mathcal{O}(1-10)$ per year for LSST. Notice that the actual rate might be dominated by a few prime targets that have lower redshift star-forming galaxies, as most of the local galaxies are rather quiescent. Furthermore, this estimation is based on the assumption that all UDGs are located at the same distance as NGC1052-DF2. The event rate is coupled with the optical depth directly through Eqn.~\ref{eqn: tau_2}, and scales linearly with $D_{L}D_{LS}/D_{S}$. Since UDGs are rather close to us, $D_{LS} \simeq D_{S}$ and the amplitude of $D_{L}$ severely affects the estimated event rate. The closer the UDGs are, the smaller the Einstein rings would appear, thus reducing the optical depth and therefore the event rate. At a cost of covering fewer background galaxies by spanning a smaller angular size in the sky, UDGs located further away would act as more effective sources of microlens to induce UDG microlensing events. Since most UDGs in \citet{Zaritksy_2023} do not have measurement in recessive velocity, we cannot derive the mean $D_{L}$ for these UDGs and optimize the estimation. 

{\it Euclid}'s all-sky surveying capability has already discovered many UDG candidates \citep{marleau2025euclidquickdatarelease} with many more anticipated, where next-generation surveys for ultra-low surface brightness galaxies like ARRAKIHS \citep{Van_Damme_2024}, and wide-field instruments like {\it Roman} will provide a more pristine view of UDGs. Furthermore, {\it Euclid} will construct the {\it Euclid} deep fields through repeated observations separated by some cadences \citep{Scaramella_2021}, providing time domain information upon inspections of individual pointings. Covering an angular area of $\sim 50\,$deg, individual deep field pointings will have depth of $\sim 24-25\,$mag, comparable with LSST but with even higher angular resolution that benefits transient detection. All these, combined with LSST's monitoring or dedicated {\it JWST} searching program, make UDG microlensing detectable within the next decade according to our predictions, and the scientific use case of UDG microlensing, as we shall show in the next section, will become relevant for a more comprehensive understanding of UDGs. 


\section{Discussion}
\label{sec: discuss}

\subsection{Caustic network formation}
\label{sec: caustic}

Our analysis assumed isolated microlenses for UDGs. In reality, the critical curves (Einstein rings) of two or more microlenses will merge together when the microlenses are sufficiently close to each other. In such cases, the source plane caustic will no longer be singular (where the light curve is gaussian-like and symmetric, Eqn~\ref{eqn: mag_vs_b}), but will be in complicated shapes of caustics where sources can enter (or leave) the caustic region and attain a sharp increase (decrease) of magnification \citep[e.g.,][]{Gaudi_2002} leaving behind asymmetric light curves. To the first order, caustic networks are only formed when the two microlenses are located at where their critical curves touch each other, until where the two microlenses are located within their own Einstein rings such that they are effectively a single point mass lens where the singular cusp is formed again.

There are three cases in which the aforementioned effect could happen: (1) the pure geometric overlap of microlenses when they are projected onto the thin lens plane; (2) the stellar multiplicity where two (or more) stars are intrinsically close to each other; and (3) the existence of strong shear. Here in this section, we investigate the expected contribution of these three different factors, and examine whether the formation of a caustic network will affect our inferred UDG microlensing rate and scientific application therein.


\subsubsection{Geometric cusps}

A simple first-order estimation for the first case of geometric overlapping could be a Monte-Carlo sampling that evaluates the probability of Einstein rings overlapping with each other\footnote{Notice that the critical curves should connect together (and form caustic networks) when the critical curves are sufficiently close to each other without necessarily having actual overlapping. That, however, is only accessible via computationally costly high-resolution ray-tracing simulations in a large field-of-view, such that we only adopt this simple approximation to estimate the level of caustic crossing, and defer a detailed analysis to future work.}, given some surface mass density. Following the previous postulation, we assume a mean stellar mass of microlenses as $0.3\,M_{\odot}$, where the Einstein radius is $\sim 50000\,R_{\odot}$ ($\equiv 0.001\,$pc) for $D_{L} = 22\,$Mpc and $D_{S} = 1300\,$Mpc (at source redshift of $\sim0.5$). We then record the number of microlenses with their Einstein ring overlapping/touching with one another when we randomly sample their position on the sky within a box with some angular area, where the number of microlenses follows the surface mass density of UDG NGC1052-DF2 ($\sim 5\, M_{\odot}/\textrm{pc}^{2}$). We found among 500 realizations of the Monte Carlo, none of the realizations have microlensing Einstein rings overlapping with each other. Since this is for a rather higher redshift case considered in our work, the Einstein rings would be smaller for a lower redshift source, and the rate of overlapping would be even lower. Hence, and therefore, cusp formation owing to pure geometric overlapping is negligible for the case of UDG microlensing when the UDG is rather nearby. Notice that this should scale linearly with the lens distance, $D_{L}$, such that the geometric overlapping could be more appreciable when the UDGs are further away from us. We shall discuss the possible contribution of neighbour microlenses through shear later in Sec.~\ref{sec: shear}.

\subsubsection{Binary cusps}

\begin{figure*}[ht!]
    \centering
    \includegraphics[width=\linewidth]{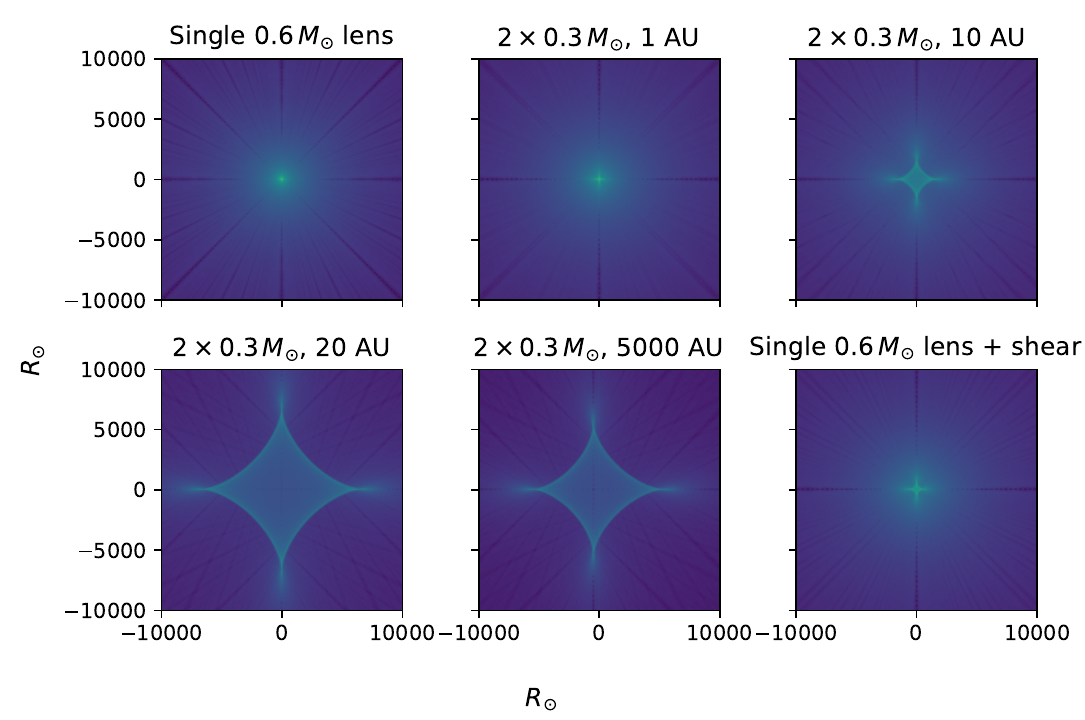}
    \caption{Source plane caustics for different UDG microlensing scenarios: (1) A single microlens with $0.6\,M_{\odot}$, (2) two microlenses of $0.3\,M_{\odot}$ separated by 1 AU, (3) two microlenses of $0.3\,M_{\odot}$ separated by 10 AU, (4) two microlenses of $0.3\,M_{\odot}$ separated by 20 AU, (5) two microlenses of $0.3\,M_{\odot}$ separated by 5000 AU (where we only show one of the cusps, as the other microlens fall far beyond the field-of-view), and (6) a single microlens with $0.6\,M_{\odot}$ alongside an external shear of $10^{-4}$. All lenses are at a distance of 20 Mpc, and a source redshift of $0.5$. The image scaling is the same for all six panels, and one can see that the maximum magnification is highest for the single microlens case. Black strips in the images are artifacts from ray tracing simulation, and do not affect our results. }
    \label{fig: caustics}
\end{figure*}

In our local universe, $\sim 40\%$ of solar-mass stars and $\sim 30\%$ of low-mass stars are known to have at least one companion star \citep[e.g.,][]{Raghavan_2010, Winters_2019}, where the typical separation between the stars \citep[$\sim \mathcal{O}(1-10)\,$AU, depending on the primary mass,][]{offner2022originevolutionmultiplestar} can be comparable with the Einstein radius we are considering for UDG microlensing.

While the caustic topology from binaries is well studied in galactic microlensing \citep[e.g.,][]{dominik1999binarygravitationallensextreme}, the typical separation of binaries is way smaller than the Einstein radius in the regime of UDG microlensing. We thus investigate when the caustic network forms in binary systems by doing ray-tracing simulations for some simple cases where we have two stars with identical mass, separated by some distances in the image plane. We consider two identical microlenses with $0.3\,M_{\odot}$ separated by different distances under the same lensing geometry of source redshift 0.5 and lens distance of $\sim 20\,$Mpc with a resolution of $\sim 20\,R_{\odot}$ in the source plane. The median binary separation distance is $\sim 10\,$AU for such low mass stars \citep{offner2022originevolutionmultiplestar}, such that we test four scenarios in our simulation with the two stars separated by $\sim 1\,$AU, $\sim 10\,$AU and $\sim 20\,$AU, as well as having a single microlens that have $0.6\,M_{\odot}$. We also show one of the two cusps \citep[as the other cusp is far-flung from the allowed field-of-view given the resolution, corresponding to the][caustic]{Chang_Refsdal_1979} for a binary case with $\sim 5000\,$AU separation, representing a very rare scenario of a low-mass wide binary.
All the caustics are shown in Fig.~\ref{fig: caustics}, where one can see that astroid caustics are forming for all cases except when there is a single microlens. Also, for close binaries, the astroid size scales with the separation squared, aligning with the literature \citep[e.g.,][]{dominik1999binarygravitationallensextreme}.


\subsubsection{Role of shear}
\label{sec: shear}

Shear is known to be capable of unfolding point-like caustics from isolated microlenses to form astroid caustics. On top of the binary lenses, we also test a case where a single microlens is exposed to a background shear of $10^{-4}$ as shown in Fig.~\ref{fig: caustics}, which is roughly the minimum value of shear that will lead to the formation of astroid caustics. Notice that this is in a resolution of $\sim 20\,R_{\odot}$, such that in lower resolution (larger source), the minimum shear strength that will lead to astroid caustic formation will be larger. There are two sources of shear that are involved in UDG microlensing: (1) intrinsic: shear coming from nearby stars, as well as more compact structures like globular clusters (GCs) inside the UDG, and the net effect from the host galaxy (including dark matter, if any), and (2)  extrinsic: external shear from large scale structure.



The shear strength for a point-mass lens scales with $(R_{E}/d)^{2}$ and $d$ the distance from the lens in the image plane. Again, the stellar density of $\sim 5\,M_{\odot}/\textrm{pc}^{2}$ considered in our case of NGC1052-DF2 is rather low, such that one can expect an average separation between microlenses with mean masses of $0.3\,M_{\odot}$ to be $\sim 0.25\,\textrm{pc}$. This means that the mean shear strength contributed by neighbour microlenses is of the order of $\sim 10^{-5}$. 
We can also refer to the case where the two identical microlenses are separated by $\sim 5000\,$AU apart, as shown in the lower middle panel in Fig.~\ref{fig: caustics}. This separation is $\sim 1/10$ of the typical separation between individual microlenses, where the expected shear strength is $\sim 0.01$. Given that the cusps barely appear in the case of shear strength of $\sim 10^{-4}$ as shown earlier in Fig.~\ref{fig: caustics}, astroid caustic formation that is larger than sources with sizes under our consideration ($\sim 10-100\,R_{\odot}$) is unlikely owing to shear from neighbour microlenses.

\begin{figure}
    \centering
    \includegraphics[width=\linewidth]{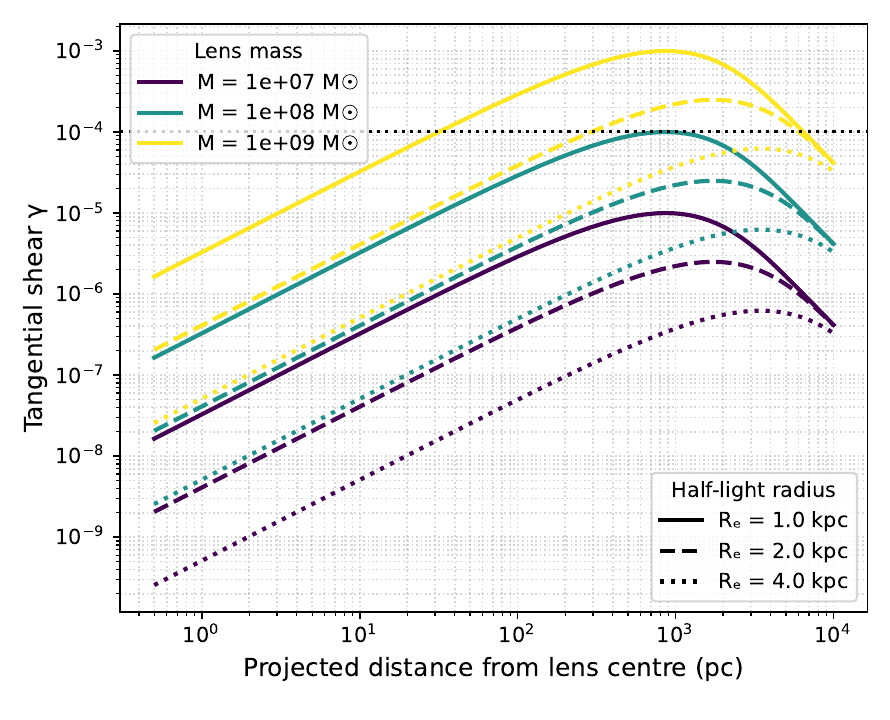}
    \caption{Tangential shear strength as a function of distance from the center of Sersic profiles, assuming a Sersic index of 1. We show the case for different stellar masses and different half-light radii reflected by the line style.}
    \label{fig: Sersic_shear}
\end{figure}

Furthermore, we can consider the collective effect of all the neighbouring stars through the stellar profile of the UDG if they are indeed lacking dark matter. UDGs are known to be well-fit by \citet{Sersic_1963} profiles with Sersic index, $n$, of $\sim 1$ owing to their diffuse nature \citep[e.g.,][]{Koda_2015, Roman_2017}. We solve for the shear strength of Sersic profiles, assuming a circularly symmetric profile following different half-light radii and stellar masses, as shown in Fig.~\ref{fig: Sersic_shear}. If $10^{-4}$ is the typical shear strength that will lead to cusp formation for individual microlenses, then Sersic profiles will only contribute sufficient shear to form astroid caustics in UDG microlenses if they are more massive than $10^{9}\,M_{\odot}$ at some distance from the center, depending on the half-light radius. For lower-mass UDGs, the mass profile alone does not contribute enough shear to form cusps that are large enough for UDG microlensing, aligning with our previous postulation when only considering the contribution from individual stars.

\begin{figure}
    \centering
    \includegraphics[width=\linewidth]{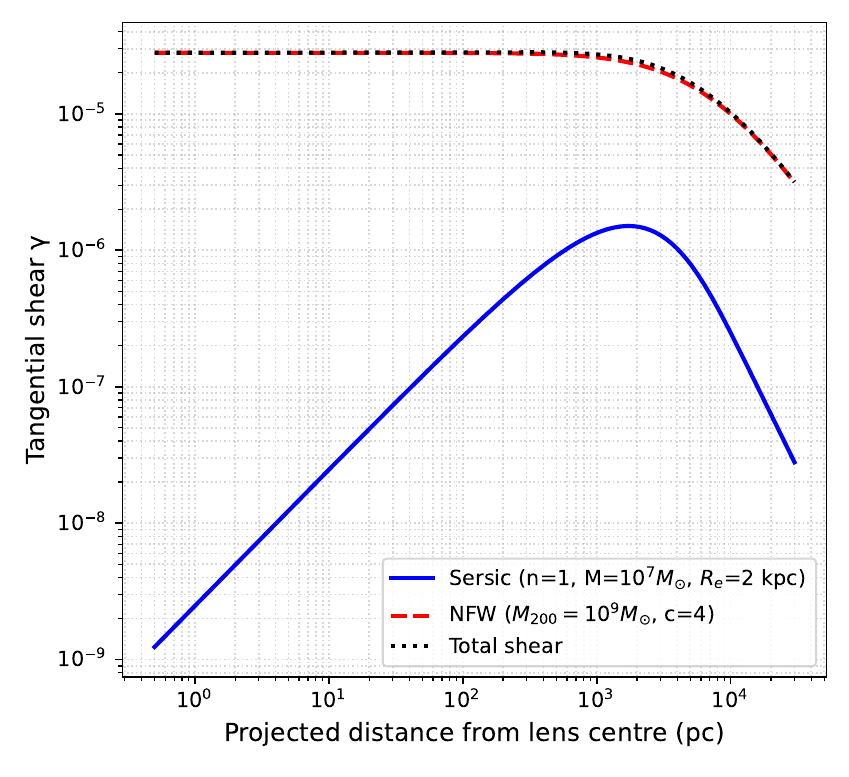}
    \caption{Shear strength as a function of projected distance, adopting the mass profile of NGC1052-DF2 approximated as the sum (black dotted line) of an NFW profile (red) and a Sersic (blue) as reported in \citet{Trujillo_2019} with a closer lens distance of $\sim13\,$Mpc. One can see that in this case, the NFW component dominates the shear. }
    \label{fig: nfw_sersic}
\end{figure}

As pointed out by \citet{Trujillo_2019}, the distance to NGC1052-DF2 could have been shorter ($\sim 13\,$Mpc) with an inferred stellar mass of $\sim 6\times10^{7}\,M_{\odot}$ and allow for a dark matter mass of $\sim 10^{9}
\,M_{\odot}$. We estimate the shear strength for such a case while revising the lens distance to $13\,$Mpc, again assuming a circularly symmetric \citet{NFW} profile with concentration of $c = 4$ following the analytical formalism in \citet{Wright_2000}, combining with a Sersic profile that represents the stellar profile as we have done previously. The resulting shear profile is shown in Fig.~\ref{fig: nfw_sersic}, where one can see that adopting the mass profile from \citet{Trujillo_2019}, combining dark matter and baryonic component, will only lead to shear strength of $\sim 10^{-5}$, which is below the level to create astroid caustics from shear. Nevertheless, for a more massive UDG with more dark matter (or a slightly different lens geometry), the contribution of dark matter (if any) would become non-negligible and would lead to cusp formation everywhere within the UDGs.

\begin{figure}
    \centering
    \includegraphics[width=\linewidth]{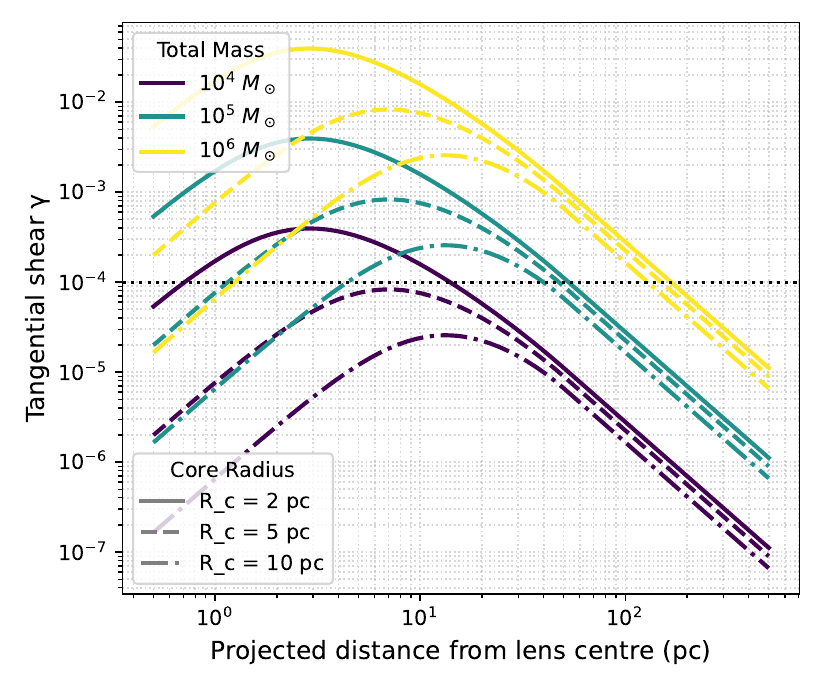}
    \caption{Tangential shear strength as a function of distance from GC, given different GC masses as reflected by different colours, and different core radii as reflected by the linestyles. We show the shear strength of $10^{-4}$ (which leads to the cusp formation in the lower right panel of Fig.~\ref{fig: caustics}) as a dotted horizontal line as reference.}
    \label{fig: GC_shear}
\end{figure}

Another contributor is the GC population, whereas UDGs can be classified as GC-rich ($>20$) and GC-poor ($<20$) \citep[see, e.g.,][]{van_Dokkum_2017, Lim_2018, Forbes_2020}. For the case of NGC1052-DF2, recent spectroscopic observation confirmed a total of 16 GCs \citep{Fahrion_2025}. To investigate the shear contributed by GCs, we adopt the \citet{King_1962} profile:

\begin{equation}
    \Sigma_{k} \propto [\frac{1}{1+(r/r_{c})^{2}} - \frac{1}{1+(r_{t}/r_{c})^{2}}]^{2},
\end{equation}

\noindent with $\Sigma_{k}$ the surface mass density of the king profile, $r_{c}$ the core radius, $r_{t}$ the tidal radius, and $r$ the distance from the center. We then calculate the tangential shear strength at the same lensing geometry of source redshift of 0.5, and lens distance of $\sim 20\,$Mpc, as a function of physical separation from these GCs, as shown in Fig.~\ref{fig: GC_shear}. As expected, the shear strength increases with the GC mass, whereas a larger core radius distances the peak of the tangential shear and suppresses the maximum shear possible. Again, if $10^{-4}$ is the typical shear strength that will lead to astroid caustic formation, then it means that the UDG microlensing event will demonstrate caustic crossing if it is within some radius of GCs, depending on the GC mass and core radius. For example, for a $10^{4}\,M_{\odot}$ GC with a core radius of $2\,$pc, this distance would be of $\sim 20\,$pc, but no caustic crossing would happen owing to the shear of GCs if these low-mass GCs have a larger core radius as the peak shear strength drops below the fiducial $\sim10^{-4}$ value. On the other hand, more massive GCs will always induce astroid caustics for UDG microlenses, regardless of the core radius, to within a distance of $\sim 50\,$pc for $10^{5}\,M_{\odot}$ GCs, and $\sim 100\,$pc for $10^{6}\,M_{\odot}$ GCs.
Assuming that the GC population in UDGs is well described by the typical log-normal GC mass function \citep[e.g.,][]{Fall_2001, Jordan_2007} with a mean of $\sim 2\times 10^{5}\,M_{\odot}$ and are distributed homogeneously across the UDG, then for the case of NGC1052-DF2, the physical area of which UDG microlensing will demonstrate caustic crossing purely due to shears from GCs would be $\sim 128,000\,\textrm{pc}^{2}$, which is $\sim 0.6\%$ of the total area of NGC1052-DF2 given a effective radius of $\sim 2.2\,$kpc \citep{van_Dokkum_2018_distance}. While this effect is negligible to the first-order globally, background galaxies and UDG microlensing events therein that are in close proximity to (massive) GCs will likely be affected by the shear of GCs.

Last but not least, the last component will be cosmic shear. Given an arbitrary part of the sky, the theoretical shear strength owing to large-scale structures fluctuates around zero with some root-mean-square (rms) value of $\gamma_{\textrm{rms}}$. Here, we do a hand-waving estimation to evaluate this ambient field shear. We calculate $\gamma_{\textrm{rms}}$ with PyCCL \citep{Chisari_2019} by modelling the line-of-sight integration of the non-linear cosmic web and scale the cumulative tidal perturbations with the geometric lensing efficiency kernel. We fix the lens at $\sim 20\,$Mpc ($z \approx 0.005$) and vary the source redshift. The expected $\gamma_{\textrm{rms}}$ is of the order of $\sim 10^{-3}-10^{-2}$ for scales of $\sim 1-100$ arcminute. This simple estimation appears to align with expectations from recent calculations estimating the contribution from cosmic filaments \citep{nikjoo2026weaklensingphotometricdensity}. Even though the value is lower for a closer source (unless the source has $z\lesssim 0.01$), $\gamma_{rms}$ will very likely exceed the ``threshold'' of $\sim 10^{-4}$. In other words, the singular caustic of UDG microlenses is almost likely to always be unfolded into astroid caustics for a source with $\sim 20\,R_{\odot}$, under random cosmic shear at any given field in the sky. Since this is the expected rms, statistically speaking, there are patches of sky where the cosmic shear strength is below the threshold of $10^{-4}$. Pinning down the shear strength to such a low value, however, is extremely difficult given the expected Poissonian noise.

\begin{figure}
    \centering
    \includegraphics[width=\linewidth]{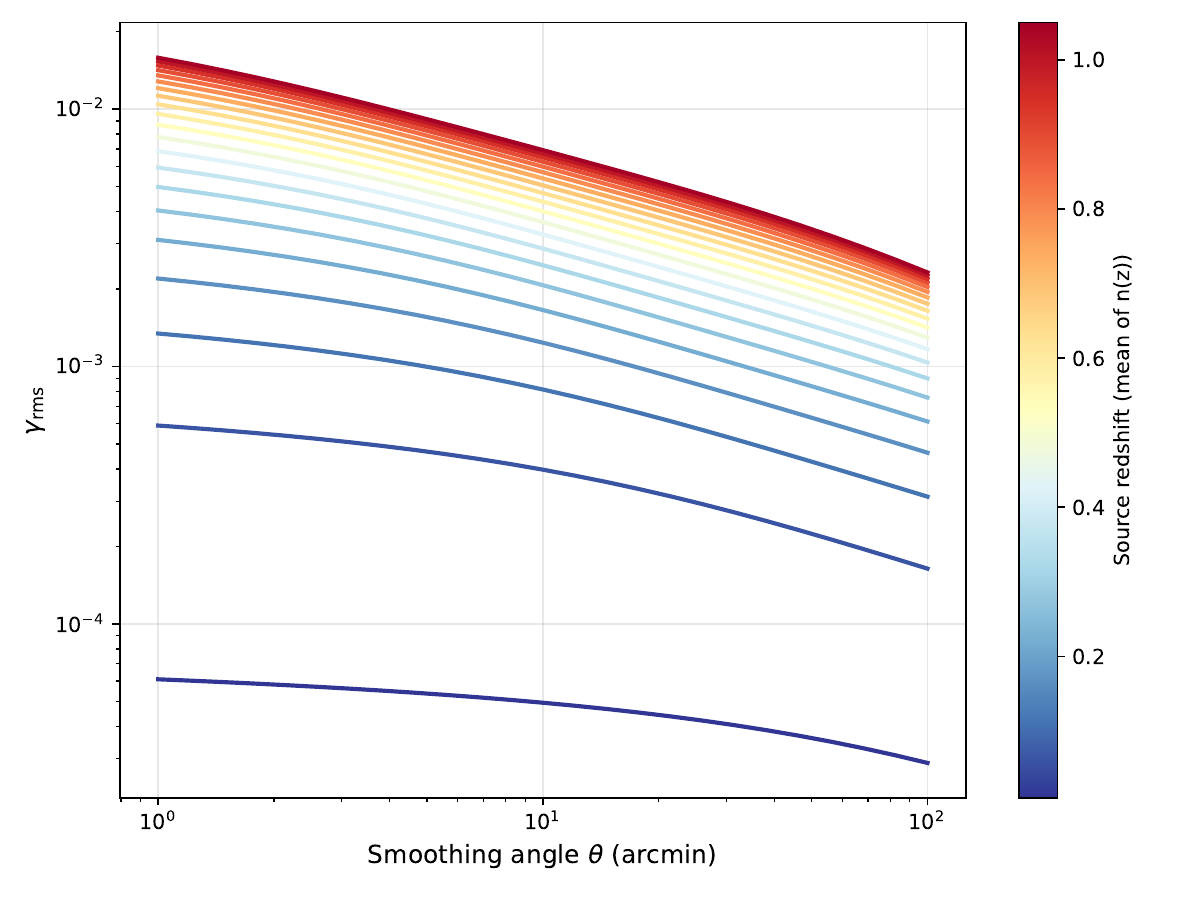}
    \caption{$\gamma_{\textrm{rms}}$ expected over different source redshifts (reflected by colour) at different angular scales as contributed by cosmic shear. }
    \label{fig: shear}
\end{figure}

%


\subsubsection{Effect on detection rate}
\label{sec: bias}

\begin{figure}
    \centering
    \includegraphics[width=\linewidth]{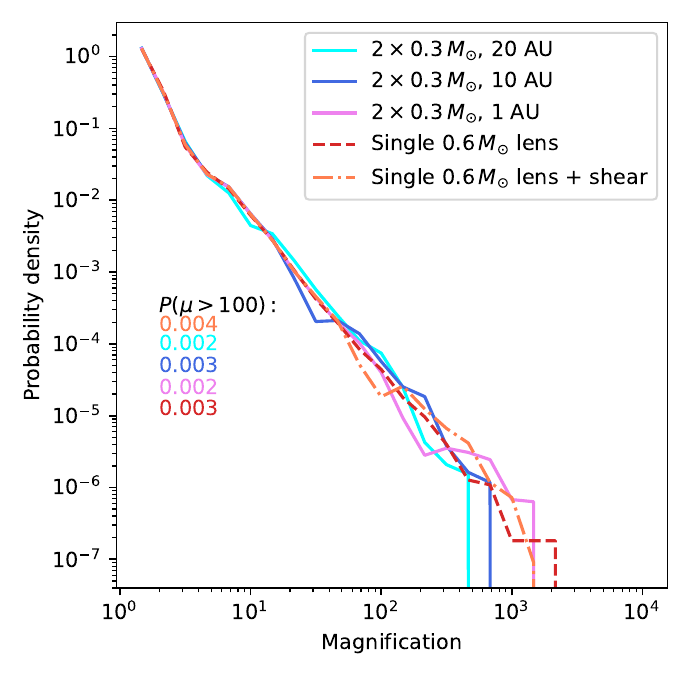}
    \caption{Magnification probability density function for all five source plane caustics shown in Fig.~\ref{fig: caustics}. We also show the probability of having magnification larger than $100$ for all these cases as texts with colour schemes following the legend.}
    \label{fig: caustic_pdfs}
\end{figure}

We can examine whether the detection of UDG microlensing events would be biased towards caustic crossing by accessing the ray-tracing simulations shown earlier in Fig.~\ref{fig: caustics} and collecting the probability distribution of magnification a background star can attain as shown in Fig.~\ref{fig: caustic_pdfs}. One can see that all five cases that share the same surface mass density within the field-of-view (i.e., except the wide-binary case) have a similar probability density function with the $\mu^{-2}$ power-law tail. Unsurprisingly, single-lens cases (alongside to that with the smallest separation) have the maximum magnification (as calculated earlier in Fig.~\ref{fig: max_mu}), and the larger the separation, the lower the maximum magnification (e.g., the cyan line that represents a binary separated by 20 AU).

As shown in Figure 5, there is a clear scaling with the size of the caustic and the separation of the binaries. Wider separations result in larger astroids, but the probability of magnification remains more or less constant, as shown in Fig.~\ref{fig: caustic_pdfs}, independent of the binary separation. This means in the case of wider binaries, the magnification has to spread over larger areas, and at the binary microcaustic, the peak magnification is smaller for the binary with 20 AU than for the binary with 10 AU, as also reflected by the probability density being truncated earlier for the case of a wider binary in Fig.~\ref{fig: caustic_pdfs}. Hence, we should expect the binary with 10 AU to produce larger magnifications during the peak, but on the other hand, the perimeter of the caustic is smaller than in the case of a binary with 20 AU separation. These two competing effects complicate the calculation of the rate. Wide binaries will magnify more background stars, but they will be magnified by smaller factors than compact binaries. 


We can adopt a fiducial minimum magnification that allows a lensed star detection through UDG microlensing as 100 magnification, to see whether the detection of UDG microlensing events is biased towards astroid caustics. Notice that this minimum magnification roughly translates to a case with a {\it JWST} threshold of $\sim 29\,$mag, where we have a source blue supergiant star with absolute magnitude of $-8$ and a radius of $\sim 20\,R_{\odot}$ that is comparable to our simulation resolution. As shown in the inset texts of Fig.~\ref{fig: caustic_pdfs}, the probability of all 5 cases having a magnification more than 100 is $\sim 0.2-0.4\%$. There is no significant trend in whether the total probability increases or decreases with the increased separation in this case. Nevertheless, as we have described in the last paragraph, wider binaries would be less powerful in terms of having a higher maximum magnification, such that they will disfavour the detection of short-duration UDG microlensing events with intrinsically dimmer stars, i.e., when the minimum magnification is raised up to $\sim 500$. Our estimated rate, assuming singular caustics, thus only provides an upper limit of the UDG microlensing event rate. 

Another possible effect of caustic crossing is that the event time-scale could have been prolonged such that the event rate as estimated through Eqn.~\ref{eqn: peryear} will be decreased. Taking the two caustic maps in the left column of Fig.~\ref{fig: caustics} as a reference, the exact same path would have a longer detection period when the astroid caustic exists: given the same detection threshold, a diagonal entry from the fold could have yielded twice the event duration, whereas an entry through the cusp would have yielded up to six times the original duration when passing through the lens center. That said, even though astroid caustic formation is very likely for UDG microlensing given our estimated shear strength, not all the microlensing events will necessarily go through the fold. For example, a trajectory can cut through the cusps without entering the caustic via the fold, such that the resulting shape of the light curve remains largely symmetric, given the signal-to-noise, like the case of regular microlensing without shear. In theory, the magnification attainable is different for the two cases, but they will be completely degenerate when one does not know the true brightness of the background star. The trajectory of the microlenses is completely stochastic, such that the estimation of how the UDG microlensing timescale will be prolonged can only be accessed via Monte-Carlo sampling.

We carry out a simple Monte-Carlo by randomly sampling trajectories in two caustic maps of (1) a single microlens with 0.6 $M_{\odot}$ (singular caustic), and (2) a binary microlens, each with 0.3 $M_{\odot}$ and separated by $10\,$AU. Among 500,000 realizations, both cases have $\sim 0.2\%$ of detecting a microlensing event if we set the threshold at $\mu_{min} = 100$, aligning with what was shown in Fig.~\ref{fig: caustic_pdfs} that there are no apparent selection biases based on caustic crossing. The mean duration of the first case is $\sim 15\pm4$ days, and is $\sim 6\pm5$ days for the second case, assuming the velocity of $\sim 300\,\textrm{km}/\textrm{s}$. Notice that there are cases where background stars have not attained sufficient magnification when they are inside the caustic, and they can be identified as two separate events when the signal-to-noise is insufficient to probe the asymmetry. In that case, the event duration will be shorter as a caustic crossing will lead to a sharper rise of magnification. The average event duration for the binary case would be longer if $\mu_{min}$ is lower, as the sources could be detected even when they are inside the caustic. Regardless, the time scale in Eqn~\ref{eqn: t_b} only differs by a factor of two in this case, and the event rate in Eqn.~\ref{eqn: peryear} remains largely unchanged even when we consider stellar multiplicity and/or external shear. As we only aimed to provide a zeroth-order event rate estimation in this paper and just to show the UDG microlensing is detectable with next-generation surveys, we do not tailor and apply the effect of caustic network formation to all the event rates as shown in Fig.~\ref{fig: detect} since it will be computationally expensive to estimate the difference for all source redshifts, and it is nevertheless still within the same order of magnitude.


\subsection{Applications}

There are three primary observables for UDG microlensing: (1) the overall event rate, (2) the spatial distribution, and (3) the light curves of events. 


\subsubsection{Event rate, surface mass density and initial mass function}

The stellar mass of a UDG is usually inferred from its luminosity via an assumed mass‑to‑light ratio, which depends critically on the low‑mass end of the IMF. For NGC1052‑DF2, dynamical measurements given an assumed mass-to-light ratio of $\sim 2$ suggest little or no dark matter \citep{van_Dokkum_2018}. However, this conclusion rests on converting luminosity into stellar mass – an IMF‑dependent step. For example, varying the IMF between a \citet{Salpeter_1955} and a \citet{Chabrier_2003} form changes the inferred stellar mass by a factor $\sim 2$. Consequently, the mass-to-light ratio could be lower under a bottom-lighter IMF, leaving room for a dark matter component.

UDG microlensing provides a direct measurement of the stellar surface mass density $\Sigma_{\star}$. The UDG microlensing event rate (Eqn.\ref{eqn: expected} and \ref{eqn: peryear}) depends on the actual number density of microlenses, which are dominated by stars with masses near the mean mass of $\hat{M}\approx 0.3 \,M_{\odot}$  – the low‑mass end of the IMF that produces most of the lensing cross section but little light. By combining the observed event rate with an independent estimate of the background star formation rate (e.g., from H$\alpha$ or H$\beta$ luminosity as we have done earlier in Sec.~\ref{sec: rate}), one can derive $\Sigma_{\star}$ and therefore infer the underlying IMF. The resulting stellar mass can then be compared with the dynamical mass from kinematics to directly reveal the dark matter fraction, free from the IMF ambiguity.

For NGC1052‑DF2 itself, the predicted event rate is too low for a near‑term measurement (Tab.~\ref{tab: table}). However, for more favourable UDGs with lower‑redshift background galaxies that are undergoing active star formation – exactly those that will be identified by Euclid \citep[e.g.,][]{marleau2025euclidquickdatarelease} – a decade of LSST monitoring, alongside dedicated {\it JWST} programs that focus on specific UDGs, will yield many events. These events will decisively test whether some UDGs truly lack dark matter or merely have a bottom‑light IMF that biases the dynamical interpretation.  Since the current measurement of stellar IMFs in UDG focuses on globular clusters \citep[e.g.,][]{Beasley_2025, Fahrion_2025}, accessing the field stars via UDG microlensing can independently constrain the IMF. Whether the globular clusters and field stars share the same IMF would also bring paramount insights into the formation channel of UDGs \citep[see review,][]{Gannon_2026}.





\subsubsection{Spatial distribution}  

The spatial distribution of event rate depends on the gradient of the surface mass density (and therefore IMF, as argued in the last paragraph) of the UDGs. Here we assumed a constant $\sim 5\, M_{\odot}\textrm{yr}^{-1}$ surface mass density to simplify our calculation. In reality, background galaxies located at the outskirts of UDGs should have a lower event rate even if they have the same SFR as galaxies that are located in the inner region of UDGs. UDG microlensing could potentially happen even a little beyond the maximum visible ``radius'' of UDGs, as such low mass stars can be sparsely populated with little associated luminosity, such that they are not distinguished from the sky background with existing observations. This provides a unique way of possibly tracing the smallest and furthest stellar population in UDGs, thus their ``true'' radii. Alternatively, an exceptionally high rate of UDG microlensing at the outskirts of UDGs could hint towards the existence of a large population of compact dark objects such as primordial black holes \citep{Carr_2021}.


\subsubsection{Light curves}

The light curve of UDG microlensing is trickier compared with the event rate because the background stars will not be detected regularly, unlike the case of local galactic microlensing. Consider a somewhat realistic case where the UDG microlensing event barely gains sufficient magnification at the peak of the light curve, such that it is detected by LSST, and {\it JWST} immediately carries out follow-up observations. Even though follow-up observations are deployed, they would be more likely to be able to capture only the second half of the light curve, losing a lot of information, and leading to severe degeneracies between velocity, lens mass, and source radii. This would be especially true for NGC1052-DF2, given that there are no very low redshift galaxies behind it. Deploying {\it JWST} for a long-duration monitoring program is extremely unrealistic, given that the event rate is rather low for galaxies further away. 

We can instead consider a more idealized case where a UDG has background sources more nearby ($z\approx 0.01-0.1$), rendering it more likely that LSST could observe microlensing events prior to the peak of the light curve. The high time cadence LSST observations of $\sim 3$ days would then potentially allow us to retrieve the full light curve (whose duration is approximated by Eqn~\ref{eqn: t_b}). In this case, since the distance to the UDG is known (although with uncertainty), as well as the distance to the source galaxy, the only variable left would be the lens mass, velocity, and the source properties, including the intrinsic brightness and the stellar radius. Photometry (six bands in LSST) or even spectroscopic measurements (by other instruments) can allow one to better understand the properties of the source star, for example, the stellar radii and intrinsic brightness, given the achromatic nature of microlensing in general. Analysis analogous to that employed for galactic microlensing can be carried out in this case to constrain the lens mass and source properties. Since such a technique is well studied in galactic microlensing \citep[e.g.,][]{berloff2025unifiedanalyticframeworkmicrolensing}, and UDG microlensing does not bring any extra insights apart from studying the stellar population in UDGs instead of local populations, we do not further detail the complications in this paper.

In an ideal case, the light curve might even allow one to probe the stellar multiplicity through the fraction of events that exhibit caustic crossing if the background shear (especially from the cosmic shear) falls below the ``threshold'' that will always lead to caustic crossing. Such a scenario could be more possible for larger source stars (with lower sensitivity towards the cusp formation), and/or for even closer background galaxies. Since geometric overlapping is unlikely (limited at $<0.2\%$ probability, as none of our 500 realizations show overlapping), when the external shear is not strong enough to create cusps means that caustic crossings are most likely dominated by binary systems in UDGs when the events are far from GCs. The full systematics, degeneracy analysis (e.g., the peculiar velocity versus shear strength and/or binary microlenses regarding the caustic crossing timescale), characterization of shear fields, and whether it is possible to decouple between shear and stellar multiplicity with a large data sample (e.g., if we only make use of larger source stars and events therein, since they have a larger tolerance to background shear) have to be explored in future work before we can further discuss the constraining power of stellar multiplicity of UDG microlensing. 

\section{Summary}

Here in this paper, we propose the idea of observing individual stars in distant galaxies via stellar microlensing with the stellar population in UDGs, with NGC1052-DF2 as a case study. Our first order approximation is consistent with the null detection so far with existing surveys like the ZTF, but indicates that detecting such UDG microlensing is viable with next-generation all-sky surveys like LSST, or with a dedicated searching program with {\it JWST}, depending on the properties of the source galaxies, including their redshifts and recent SFR. While NGC1052-DF2 is not a prime target to look for UDG microlensing events, given its lack of low-redshift background galaxies and therefore low expected LSST and {\it JWST} event rate, {\it Euclid}'s all-sky survey is expected to reveal many suitable UDGs that favour UDG microlensing events that are observable by LSST and {\it JWST}. A zeroth-order estimated total event rate of $\mathcal{O}(1-10)\,\textrm{yr}^{-1}$ is expected for LSST's all-sky monitoring.

We have also explored the contribution of the shear field towards cusp formation for UDG microlensing, finding that the external shear would play a significant role in affecting the shapes of the light curves. While the formation of cusps (from either binary microlenses or external shears) does not affect the event rate, future work is required to explore the full astrophysical potential of light curves of UDG microlensing.
Regardless, we postulate that the event rate of UDG microlensing would provide a unique opportunity in studying stellar populations in UDGs, such as the true stellar mass and the therefore low-mass end of the stellar IMF, as well as constraining the abundance of compact dark matter. Analysis of UDG microlensing events alongside other observations can potentially improve our understanding of UDG properties as well as their formation mechanism.

\begin{acknowledgements}

We thank Leo W.H. Fung for his useful discussion regarding the shear for UDG microlensing. S.K.L. acknowledges the generous support from the Croucher Foundation through the Croucher Postdoctoral Fellowship.
S.K.L., J.L., and J.N. acknowledge support from the Research Grants Council (RGC) of Hong Kong through the General Research Fund (GRF) 17302023.
T.B. acknowledges the Spanish Grant PID2023-149016NB-I00 (MINECO/AEI/FEDER, UE).
T.B. and J.Z. acknowledge the CEX2024-001491-S grant, funded by MICIU/AEI/10.13039/501100011033. 
J.M.D. acknowledges support from project PID2022-138896NB-C51 (MCIU/AEI/MINECO/FEDER, UE) Ministerio de Ciencia, Investigaci\'on y Universidades.
J.M.P. acknowledges financial support from the Complementary Plan in Astrophysics and High-Energy Physics (CA25944), project C17.I02.P02.S01.S03 CSIC, supported by the Next Generation EU funds, RRF and PRTR mechanisms, and the Government of the Autonomous Community of Cantabria.
J.N. also acknowledges the support of the Dissertation Year Fellowship issued by the University of Hong Kong.

This research is based on observations made with the NASA/ESA {\it James-Webb Space Telescope} obtained from the Space Telescope Science Institute, which is operated by the Association of Universities for Research in Astronomy, Inc., under NASA contract NAS 5-26555. These observations are associated with program GO-3990. 
The data are available at MAST: 10.17909/ms8d-t448.

We acknowledge the usage of the following programs: Astropy \citep{astropy:2013, astropy:2018, astropy:2022}, Numpy \citep{numpy}, Matplotlib \citep{matplotlib}, and Scipy \citep{Scipy}.
\end{acknowledgements}

%
\bibliographystyle{aa} 
\bibliography{bib} 

@misc{palencia2026statisticalstudy100magnified,
      title={First Statistical Study of Over 100 Magnified Stellar Events at Redshift $z \approx 0.725$ with JWST}, 
      author={J. M. Palencia and Fengwu Sun and J. M. Diego and Yoshinobu Fudamoto and Anton M. Koekemoer and Christopher N. A. Willmer and Eduardo Iani and Xiaojing Lin and Justin D. R. Pierel and Alfred Amruth and Tom Broadhurst and W. Chen and Liang Dai and Daniel Espada and Alexei V. Filippenko and Seiji Fujimoto and Patrick L. Kelly and Mingyu Li and Sung Kei Li and Ashish Kumar Meena and Jordi Miralda-Escudé and P. Morilla and Mitchell F. Struble and Hayley Williams and Rogier A. Windhorst and E. Zackrisson and Ruwen Zhou and Adi Zitrin},
      year={2026},
      eprint={2604.22702},
      archivePrefix={arXiv},
      primaryClass={astro-ph.GA},
      url={https://arxiv.org/abs/2604.22702}, 
}

@article{Liebes_1964,
  title = {Gravitational Lenses},
  author = {Liebes, Sidney},
  journal = {Phys. Rev.},
  volume = {133},
  issue = {3B},
  pages = {B835--B844},
  numpages = {0},
  year = {1964},
  month = {Feb},
  publisher = {American Physical Society},
  doi = {10.1103/PhysRev.133.B835},
  url = {https://link.aps.org/doi/10.1103/PhysRev.133.B835}
}

@article{Haslbauer_2023,
   title={The cosmological star formation history from the Local Cosmological Volume of galaxies and constraints on the matter homogeneity},
   volume={524},
   ISSN={1365-2966},
   url={http://dx.doi.org/10.1093/mnras/stad1986},
   DOI={10.1093/mnras/stad1986},
   number={3},
   journal={Monthly Notices of the Royal Astronomical Society},
   publisher={Oxford University Press (OUP)},
   author={Haslbauer, Moritz and Kroupa, Pavel and Jerabkova, Tereza},
   year={2023},
   month=jul, pages={3252–3262} }

@article{Ma_2011,
  title = {Peculiar velocity field: Constraining the tilt of the Universe},
  author = {Ma, Yin-Zhe and Gordon, Christopher and Feldman, Hume A.},
  journal = {Phys. Rev. D},
  volume = {83},
  issue = {10},
  pages = {103002},
  numpages = {7},
  year = {2011},
  month = {May},
  publisher = {American Physical Society},
  doi = {10.1103/PhysRevD.83.103002},
  url = {https://link.aps.org/doi/10.1103/PhysRevD.83.103002}
}

@article{Hudson_2004,
    author = {Hudson, Michael J. and Smith, Russell J. and Lucey, John R. and Branchini, Enzo},
    title = {Streaming motions of galaxy clusters within 12 000 km s−1– V. The peculiar velocity field},
    journal = {Monthly Notices of the Royal Astronomical Society},
    volume = {352},
    number = {1},
    pages = {61-75},
    year = {2004},
    month = {07},
    issn = {0035-8711},
    doi = {10.1111/j.1365-2966.2004.07893.x},
    url = {https://doi.org/10.1111/j.1365-2966.2004.07893.x},
    eprint = {https://academic.oup.com/mnras/article-pdf/352/1/61/3197555/352-1-61.pdf},
}

@article{Lee_2024,
doi = {10.3847/1538-4357/ad2932},
url = {https://doi.org/10.3847/1538-4357/ad2932},
year = {2024},
month = {apr},
publisher = {The American Astronomical Society},
volume = {966},
number = {1},
pages = {72},
author = {Lee, Joohyun and Shin, Eun-jin and Kim, Ji-hoon and Shapiro, Paul R. and Chung, Eunwoo},
title = {Multiple Beads on a String: Dark-matter-deficient Galaxy Formation in a Mini-Bullet Satellite?atellite Galaxy Collision},
journal = {The Astrophysical Journal}
}

@article{Silk_2019,
    author = {Silk, Joseph},
    title = {Ultra-diffuse galaxies without dark matter},
    journal = {Monthly Notices of the Royal Astronomical Society: Letters},
    volume = {488},
    number = {1},
    pages = {L24-L28},
    year = {2019},
    month = {07},
    issn = {1745-3925},
    doi = {10.1093/mnrasl/slz090},
    url = {https://doi.org/10.1093/mnrasl/slz090},
    eprint = {https://academic.oup.com/mnrasl/article-pdf/488/1/L24/56978173/mnrasl_488_1_l24.pdf},
}

@misc{keim2026galaxymissingdarkmatter,
      title={A Third Galaxy Missing Dark Matter along a Trail of Galaxies in the NGC 1052 Field}, 
      author={Michael A. Keim and Pieter van Dokkum and Zili Shen and Shany Danieli and Imad Pasha},
      year={2026},
      eprint={2603.15860},
      archivePrefix={arXiv},
      primaryClass={astro-ph.GA},
      url={https://arxiv.org/abs/2603.15860}, 
}

@ARTICLE{Li_2025_BRratio,
       author = {{Li}, Sung Kei and {Palencia}, Jose M. and {Diego}, Jose M. and {Lim}, Jeremy and {Kelly}, Patrick L. and {Meena}, Ashish K. and {Nianias}, James and {Williams}, Hayley and {Williams}, Liliya L.~R. and {Zitrin}, Adi and {Broadhurst}, Thomas J.},
        title = "{Transient star blue-to-red ratio and star formation history in z {\ensuremath{\gtrsim}} 1 lensed galaxies}",
      journal = {\aap},
         year = 2025,
        month = dec,
       volume = {704},
          eid = {A309},
        pages = {A309},
          doi = {10.1051/0004-6361/202556037},
archivePrefix = {arXiv},
       eprint = {2506.17565},
 primaryClass = {astro-ph.CO},
       adsurl = {https://ui.adsabs.harvard.edu/abs/2025A&A...704A.309L}
}

@article{Yang_2020,
   title={Self-Interacting Dark Matter and the Origin of Ultradiffuse Galaxies NGC1052-DF2 and -DF4},
   volume={125},
   ISSN={1079-7114},
   url={http://dx.doi.org/10.1103/PhysRevLett.125.111105},
   DOI={10.1103/physrevlett.125.111105},
   number={11},
   journal={Physical Review Letters},
   publisher={American Physical Society (APS)},
   author={Yang, Daneng and Yu, Hai-Bo and An, Haipeng},
   year={2020},
   month=sep }

@article{Pozo_2021,
   title={Wave dark matter and ultra-diffuse galaxies},
   volume={504},
   ISSN={1365-2966},
   url={http://dx.doi.org/10.1093/mnras/stab855},
   DOI={10.1093/mnras/stab855},
   number={2},
   journal={Monthly Notices of the Royal Astronomical Society},
   publisher={Oxford University Press (OUP)},
   author={Pozo, Alvaro and Broadhurst, Tom and de Martino, Ivan and Luu, Hoang Nhan and Smoot, George F and Lim, Jeremy and Neyrinck, Mark},
   year={2021},
   month=mar, pages={2868–2876} }

@INPROCEEDINGS{Van_Damme_2024,
       author = {{Corral van Damme}, C. and {Prod'Homme}, T. and {Isaak}, K. and {R{\"u}hl}, T. and {Sirianni}, M.},
        title = "{ARRAKIHS: ESA's new fast-implementation science mission}",
    booktitle = {Space Telescopes and Instrumentation 2024: Optical, Infrared, and Millimeter Wave},
         year = 2024,
       editor = {{Coyle}, Laura E. and {Matsuura}, Shuji and {Perrin}, Marshall D.},
       series = {Society of Photo-Optical Instrumentation Engineers (SPIE) Conference Series},
       volume = {13092},
        month = aug,
          eid = {130920Q},
        pages = {130920Q},
          doi = {10.1117/12.3020186},
       adsurl = {https://ui.adsabs.harvard.edu/abs/2024SPIE13092E..0QC}
}

@ARTICLE{Zaritksy_2023,
       author = {{Zaritsky}, Dennis and {Donnerstein}, Richard and {Dey}, Arjun and {Karunakaran}, Ananthan and {Kadowaki}, Jennifer and {Khim}, Donghyeon J. and {Spekkens}, Kristine and {Zhang}, Huanian},
        title = "{Systematically Measuring Ultra-diffuse Galaxies (SMUDGes). V. The Complete SMUDGes Catalog and the Nature of Ultradiffuse Galaxies}",
      journal = {\apjs},
         year = 2023,
        month = aug,
       volume = {267},
       number = {2},
          eid = {27},
        pages = {27},
          doi = {10.3847/1538-4365/acdd71},
archivePrefix = {arXiv},
       eprint = {2306.01524},
 primaryClass = {astro-ph.GA},
       adsurl = {https://ui.adsabs.harvard.edu/abs/2023ApJS..267...27Z}
}

@ARTICLE{Kennicutt_1983,
       author = {{Kennicutt}, Jr., R.~C. and {Kent}, S.~M.},
        title = "{A survey of H-alpha emission in normal galaxies.}",
      journal = {\aj},
         year = 1983,
        month = aug,
       volume = {88},
        pages = {1094-1107},
          doi = {10.1086/113399},
       adsurl = {https://ui.adsabs.harvard.edu/abs/1983AJ.....88.1094K}
}

@ARTICLE{Wyrzykowski_2011,
       author = {{Wyrzykowski}, {\L}. and {Koz{\l}owski}, S. and {Skowron}, J. and {Udalski}, A. and {Szyma{\'n}ski}, M.~K. and {Kubiak}, M. and {Pietrzy{\'n}ski}, G. and {Soszy{\'n}ski}, I. and {Szewczyk}, O. and {Ulaczyk}, K. and {Poleski}, R.},
        title = "{The OGLE view of microlensing towards the Magellanic Clouds - III. Ruling out subsolar MACHOs with the OGLE-III LMC data}",
      journal = {\mnras},
         year = 2011,
        month = may,
       volume = {413},
       number = {1},
        pages = {493-508},
          doi = {10.1111/j.1365-2966.2010.18150.x},
archivePrefix = {arXiv},
       eprint = {1012.1154},
 primaryClass = {astro-ph.GA},
       adsurl = {https://ui.adsabs.harvard.edu/abs/2011MNRAS.413..493W}
}

@ARTICLE{Moniez_2010,
       author = {{Moniez}, Marc},
        title = "{Microlensing as a probe of the Galactic structure: 20 years of microlensing optical depth studies}",
      journal = {General Relativity and Gravitation},
         year = 2010,
        month = sep,
       volume = {42},
       number = {9},
        pages = {2047-2074},
          doi = {10.1007/s10714-009-0925-4},
archivePrefix = {arXiv},
       eprint = {1001.2707},
 primaryClass = {astro-ph.GA},
       adsurl = {https://ui.adsabs.harvard.edu/abs/2010GReGr..42.2047M}
}

@ARTICLE{Tsapras_2018,
       author = {{Tsapras}, Yiannis},
        title = "{Microlensing Searches for Exoplanets}",
      journal = {Geosciences},
         year = 2018,
        month = sep,
       volume = {8},
       number = {10},
          eid = {365},
        pages = {365},
          doi = {10.3390/geosciences8100365},
archivePrefix = {arXiv},
       eprint = {1810.02691},
 primaryClass = {astro-ph.EP},
       adsurl = {https://ui.adsabs.harvard.edu/abs/2018Geosc...8..365T}
}

@ARTICLE{Afonso_1999,
       author = {{Afonso}, C. and {Alard}, C. and {Albert}, J.~N. and {Andersen}, J. and {Ansari}, R. and {Aubourg}, {\'E}. and {Bareyre}, P. and {Bauer}, F. and {Beaulieu}, J.~P. and {Bouquet}, A. and {Char}, S. and {Charlot}, X. and {Couchot}, F. and {Coutures}, C. and {Derue}, F. and {Ferlet}, R. and {Glicenstein}, J.~F. and {Goldman}, B. and {Gould}, A. and {Graff}, D. and {Gros}, M. and {Haissinski}, J. and {Hamilton}, J.~C. and {Hardin}, D. and {de Kat}, J. and {Kim}, A. and {Lasserre}, T. and {Lesquoy}, {\'E}. and {Loup}, C. and {Magneville}, C. and {Mansoux}, B. and {Marquette}, J.~B. and {Maurice}, {\'E}. and {Milsztajn}, A. and {Moniez}, M. and {Palanque-Delabrouille}, N. and {Perdereau}, O. and {Pr{\'e}vot}, L. and {Regnault}, N. and {Rich}, J. and {Spiro}, M. and {Vidal-Madjar}, A. and {Vigroux}, L. and {Zylberajch}, S. and {EROS Collaboration}},
        title = "{Microlensing towards the Small Magellanic Cloud EROS 2 two-year analysis}",
      journal = {\aap},
         year = 1999,
        month = apr,
       volume = {344},
        pages = {L63-L66},
       adsurl = {https://ui.adsabs.harvard.edu/abs/1999A&A...344L..63A}
}

@ARTICLE{Niikura_2019,
       author = {{Niikura}, Hiroko and {Takada}, Masahiro and {Yasuda}, Naoki and {Lupton}, Robert H. and {Sumi}, Takahiro and {More}, Surhud and {Kurita}, Toshiki and {Sugiyama}, Sunao and {More}, Anupreeta and {Oguri}, Masamune and {Chiba}, Masashi},
        title = "{Microlensing constraints on primordial black holes with Subaru/HSC Andromeda observations}",
      journal = {Nature Astronomy},
         year = 2019,
        month = apr,
       volume = {3},
        pages = {524-534},
          doi = {10.1038/s41550-019-0723-1},
archivePrefix = {arXiv},
       eprint = {1701.02151},
 primaryClass = {astro-ph.CO},
       adsurl = {https://ui.adsabs.harvard.edu/abs/2019NatAs...3..524N}
}

@misc{offner2022originevolutionmultiplestar,
      title={The Origin and Evolution of Multiple Star Systems}, 
      author={Stella S. R. Offner and Maxwell Moe and Kaitlin M. Kratter and Sarah I. Sadavoy and Eric L. N. Jensen and John J. Tobin},
      year={2022},
      eprint={2203.10066},
      archivePrefix={arXiv},
      primaryClass={astro-ph.SR},
      url={https://arxiv.org/abs/2203.10066}, 
}

@ARTICLE{Gaudi_2002,
       author = {{Gaudi}, B. Scott and {Petters}, A.~O.},
        title = "{Gravitational Microlensing near Caustics. I. Folds}",
      journal = {\apj},
         year = 2002,
        month = aug,
       volume = {574},
       number = {2},
        pages = {970-984},
          doi = {10.1086/341063},
archivePrefix = {arXiv},
       eprint = {astro-ph/0112531},
 primaryClass = {astro-ph},
       adsurl = {https://ui.adsabs.harvard.edu/abs/2002ApJ...574..970G}
}

@misc{berloff2025unifiedanalyticframeworkmicrolensing,
      title={A Unified Analytic Framework for Microlensing Caustics: Geode Solutions and Hyper--Catalan Signatures}, 
      author={Gleb Berloff and Natalia G. Berloff},
      year={2025},
      eprint={2511.15756},
      archivePrefix={arXiv},
      primaryClass={astro-ph.IM},
      url={https://arxiv.org/abs/2511.15756}, 
}

@misc{dominik1999binarygravitationallensextreme,
      title={The binary gravitational lens and its extreme cases}, 
      author={M. Dominik},
      year={1999},
      eprint={astro-ph/9903014},
      archivePrefix={arXiv},
      primaryClass={astro-ph},
      url={https://arxiv.org/abs/astro-ph/9903014}, 
}

@article{Lim_2018,
doi = {10.3847/1538-4357/aacb81},
url = {https://doi.org/10.3847/1538-4357/aacb81},
year = {2018},
month = {jul},
publisher = {The American Astronomical Society},
volume = {862},
number = {1},
pages = {82},
author = {Lim, Sungsoon and Peng, Eric W. and Côté, Patrick and Sales, Laura V. and Brok, Mark den and Blakeslee, John P. and Guhathakurta, Puragra},
title = {The Globular Cluster Systems of Ultra-diffuse Galaxies in the Coma Cluster},
journal = {The Astrophysical Journal}
}

@ARTICLE{King_1962,
       author = {{King}, Ivan},
        title = "{The structure of star clusters. I. an empirical density law}",
      journal = {\aj},
         year = 1962,
        month = oct,
       volume = {67},
        pages = {471},
          doi = {10.1086/108756},
       adsurl = {https://ui.adsabs.harvard.edu/abs/1962AJ.....67..471K}
}

@ARTICLE{Jordan_2007,
       author = {{Jord{\'a}n}, Andr{\'e}s and {McLaughlin}, Dean E. and {C{\^o}t{\'e}}, Patrick and {Ferrarese}, Laura and {Peng}, Eric W. and {Mei}, Simona and {Villegas}, Daniela and {Merritt}, David and {Tonry}, John L. and {West}, Michael J.},
        title = "{The ACS Virgo Cluster Survey. XII. The Luminosity Function of Globular Clusters in Early-Type Galaxies}",
      journal = {\apjs},
         year = 2007,
        month = jul,
       volume = {171},
       number = {1},
        pages = {101-145},
          doi = {10.1086/516840},
archivePrefix = {arXiv},
       eprint = {astro-ph/0702496},
 primaryClass = {astro-ph},
       adsurl = {https://ui.adsabs.harvard.edu/abs/2007ApJS..171..101J}
}

@ARTICLE{Fall_2001,
       author = {{Fall}, S. Michael and {Zhang}, Qing},
        title = "{Dynamical Evolution of the Mass Function of Globular Star Clusters}",
      journal = {\apj},
         year = 2001,
        month = nov,
       volume = {561},
       number = {2},
        pages = {751-765},
          doi = {10.1086/323358},
archivePrefix = {arXiv},
       eprint = {astro-ph/0107298},
 primaryClass = {astro-ph},
       adsurl = {https://ui.adsabs.harvard.edu/abs/2001ApJ...561..751F}
}

@article{Forbes_2020,
   title={Globular clusters in Coma cluster ultra-diffuse galaxies (UDGs): evidence for two types of UDG?},
   volume={492},
   ISSN={1365-2966},
   url={http://dx.doi.org/10.1093/mnras/staa180},
   DOI={10.1093/mnras/staa180},
   number={4},
   journal={Monthly Notices of the Royal Astronomical Society},
   publisher={Oxford University Press (OUP)},
   author={Forbes, Duncan A and Alabi, Adebusola and Romanowsky, Aaron J and Brodie, Jean P and Arimoto, Nobuo},
   year={2020},
   month=Jan, pages={4874–4883} }

@article{van_Dokkum_2017,
doi = {10.3847/2041-8213/aa7ca2},
url = {https://doi.org/10.3847/2041-8213/aa7ca2},
year = {2017},
month = {jul},
publisher = {The American Astronomical Society},
volume = {844},
number = {1},
pages = {L11},
author = {van Dokkum, Pieter and Abraham, Roberto and Romanowsky, Aaron J. and Brodie, Jean and Conroy, Charlie and Danieli, Shany and Lokhorst, Deborah and Merritt, Allison and Mowla, Lamiya and Zhang, Jielai},
title = {Extensive Globular Cluster Systems Associated with Ultra Diffuse Galaxies in the Coma Cluster},
journal = {The Astrophysical Journal Letters}
}

@article{Abe_2004,
   title={Search for Low-Mass Exoplanets by Gravitational Microlensing at High Magnification},
   volume={305},
   ISSN={1095-9203},
   url={http://dx.doi.org/10.1126/science.1100714},
   DOI={10.1126/science.1100714},
   number={5688},
   journal={Science},
   publisher={American Association for the Advancement of Science (AAAS)},
   author={Abe, F. and Bennett, D. P. and Bond, I. A. and Eguchi, S. and Furuta, Y. and Hearnshaw, J. B. and Kamiya, K. and Kilmartin, P. M. and Kurata, Y. and Masuda, K. and Matsubara, Y. and Muraki, Y. and Noda, S. and Okajima, K. and Rakich, A. and Rattenbury, N. J. and Sako, T. and Sekiguchi, T. and Sullivan, D. J. and Sumi, T. and Tristram, P. J. and Yanagisawa, T. and Yock, P. C. M. and Gal-Yam, A. and Lipkin, Y. and Maoz, D. and Ofek, E. O. and Udalski, A. and Szewczyk, O. and ZÌebrunÌ, K. and SoszynÌski, I. and SzymanÌski, M. K. and Kubiak, M. and PietrzynÌski, G. and Wyrzykowski, L.},
   year={2004},
   month=aug, pages={1264–1266} }

@inbook{Mroz_2024_exo,
   title={Exoplanet Occurrence Rates from Microlensing Surveys},
   ISBN={9783319306483},
   url={http://dx.doi.org/10.1007/978-3-319-30648-3_208-1},
   DOI={10.1007/978-3-319-30648-3_208-1},
   booktitle={Handbook of Exoplanets},
   publisher={Springer International Publishing},
   author={Mróz, Przemek and Poleski, Radosław},
   year={2024},
   pages={1–23} }

@article{Rodriguez_2022,
doi = {10.3847/1538-4357/ac51cc},
url = {https://doi.org/10.3847/1538-4357/ac51cc},
year = {2022},
month = {mar},
publisher = {The American Astronomical Society},
volume = {927},
number = {2},
pages = {150},
author = {Rodriguez, Antonio C. and Mróz, Przemek and Kulkarni, Shrinivas R. and Andreoni, Igor and Bellm, Eric C. and Dekany, Richard and Drake, Andrew J. and Duev, Dmitry A. and Graham, Matthew J. and Masci, Frank J. and Prince, Thomas A. and Riddle, Reed and Shupe, David L.},
title = {Microlensing Events in the Galactic Plane Using the Zwicky Transient Facility},
journal = {The Astrophysical Journal}
}

@article{Mroz_2019,
doi = {10.3847/1538-4365/ab426b},
url = {https://doi.org/10.3847/1538-4365/ab426b},
year = {2019},
month = {oct},
publisher = {The American Astronomical Society},
volume = {244},
number = {2},
pages = {29},
author = {Mróz, Przemek and Udalski, Andrzej and Skowron, Jan and Szymański, Michał K. and Soszyński, Igor and Wyrzykowski, Łukasz and Pietrukowicz, Paweł and Kozłowski, Szymon and Poleski, Radosław and Ulaczyk, Krzysztof and Rybicki, Krzysztof and Iwanek, Patryk},
title = {Microlensing Optical Depth and Event Rate toward the Galactic Bulge from 8 yr of OGLE-IV Observations},
journal = {The Astrophysical Journal Supplement Series}
}

@article{Niikura_2019_PBH,
   title={Constraints on Earth-mass primordial black holes from OGLE 5-year microlensing events},
   volume={99},
   ISSN={2470-0029},
   url={http://dx.doi.org/10.1103/PhysRevD.99.083503},
   DOI={10.1103/physrevd.99.083503},
   number={8},
   journal={Physical Review D},
   publisher={American Physical Society (APS)},
   author={Niikura, Hiroko and Takada, Masahiro and Yokoyama, Shuichiro and Sumi, Takahiro and Masaki, Shogo},
   year={2019},
   month=apr }

@ARTICLE{Kelly_2018_Icarus,
       author = {{Kelly}, Patrick L. and {Diego}, Jose M. and {Rodney}, Steven and {Kaiser}, Nick and {Broadhurst}, Tom and {Zitrin}, Adi and {Treu}, Tommaso and {P{\'e}rez-Gonz{\'a}lez}, Pablo G. and {Morishita}, Takahiro and {Jauzac}, Mathilde and {Selsing}, Jonatan and {Oguri}, Masamune and {Pueyo}, Laurent and {Ross}, Timothy W. and {Filippenko}, Alexei V. and {Smith}, Nathan and {Hjorth}, Jens and {Cenko}, S. Bradley and {Wang}, Xin and {Howell}, D. Andrew and {Richard}, Johan and {Frye}, Brenda L. and {Jha}, Saurabh W. and {Foley}, Ryan J. and {Norman}, Colin and {Bradac}, Marusa and {Zheng}, Weikang and {Brammer}, Gabriel and {Benito}, Alberto Molino and {Cava}, Antonio and {Christensen}, Lise and {de Mink}, Selma E. and {Graur}, Or and {Grillo}, Claudio and {Kawamata}, Ryota and {Kneib}, Jean-Paul and {Matheson}, Thomas and {McCully}, Curtis and {Nonino}, Mario and {P{\'e}rez-Fournon}, Ismael and {Riess}, Adam G. and {Rosati}, Piero and {Schmidt}, Kasper Borello and {Sharon}, Keren and {Weiner}, Benjamin J.},
        title = "{Extreme magnification of an individual star at redshift 1.5 by a galaxy-cluster lens}",
      journal = {Nature Astronomy},
         year = 2018,
        month = apr,
       volume = {2},
        pages = {334-342},
          doi = {10.1038/s41550-018-0430-3},
archivePrefix = {arXiv},
       eprint = {1706.10279},
 primaryClass = {astro-ph.GA},
       adsurl = {https://ui.adsabs.harvard.edu/abs/2018NatAs...2..334K}
}

@ARTICLE{Kennicutt_1998,
       author = {{Kennicutt}, Jr., Robert C.},
        title = "{Star Formation in Galaxies Along the Hubble Sequence}",
      journal = {\araa},
         year = 1998,
        month = jan,
       volume = {36},
        pages = {189-232},
          doi = {10.1146/annurev.astro.36.1.189},
archivePrefix = {arXiv},
       eprint = {astro-ph/9807187},
 primaryClass = {astro-ph},
       adsurl = {https://ui.adsabs.harvard.edu/abs/1998ARA&A..36..189K}
}

@ARTICLE{Kennicutt_2012,
       author = {{Kennicutt}, Robert C. and {Evans}, Neal J.},
        title = "{Star Formation in the Milky Way and Nearby Galaxies}",
      journal = {\araa},
         year = 2012,
        month = sep,
       volume = {50},
        pages = {531-608},
          doi = {10.1146/annurev-astro-081811-125610},
archivePrefix = {arXiv},
       eprint = {1204.3552},
 primaryClass = {astro-ph.GA},
       adsurl = {https://ui.adsabs.harvard.edu/abs/2012ARA&A..50..531K}
}

@article{astropy:2013,
Adsurl = {http://adsabs.harvard.edu/abs/2013A%26A...558A..33A},
Archiveprefix = {arXiv},
Author = {{Astropy Collaboration} and {Robitaille}, T.~P. and {Tollerud}, E.~J. and {Greenfield}, P. and {Droettboom}, M. and {Bray}, E. and {Aldcroft}, T. and {Davis}, M. and {Ginsburg}, A. and {Price-Whelan}, A.~M. and {Kerzendorf}, W.~E. and {Conley}, A. and {Crighton}, N. and {Barbary}, K. and {Muna}, D. and {Ferguson}, H. and {Grollier}, F. and {Parikh}, M.~M. and {Nair}, P.~H. and {Unther}, H.~M. and {Deil}, C. and {Woillez}, J. and {Conseil}, S. and {Kramer}, R. and {Turner}, J.~E.~H. and {Singer}, L. and {Fox}, R. and {Weaver}, B.~A. and {Zabalza}, V. and {Edwards}, Z.~I. and {Azalee Bostroem}, K. and {Burke}, D.~J. and {Casey}, A.~R. and {Crawford}, S.~M. and {Dencheva}, N. and {Ely}, J. and {Jenness}, T. and {Labrie}, K. and {Lim}, P.~L. and {Pierfederici}, F. and {Pontzen}, A. and {Ptak}, A. and {Refsdal}, B. and {Servillat}, M. and {Streicher}, O.},
Doi = {10.1051/0004-6361/201322068},
Eid = {A33},
Eprint = {1307.6212},
Journal = {\aap},
Month = oct,
Pages = {A33},
Primaryclass = {astro-ph.IM},
Title = {{Astropy: A community Python package for astronomy}},
Volume = 558,
Year = 2013}

@ARTICLE{astropy:2018,
       author = {{Astropy Collaboration} and {Price-Whelan}, A.~M. and
         {Sip{\H{o}}cz}, B.~M. and {G{\"u}nther}, H.~M. and {Lim}, P.~L. and
         {Crawford}, S.~M. and {Conseil}, S. and {Shupe}, D.~L. and
         {Craig}, M.~W. and {Dencheva}, N. and {Ginsburg}, A. and {Vand
        erPlas}, J.~T. and {Bradley}, L.~D. and {P{\'e}rez-Su{\'a}rez}, D. and
         {de Val-Borro}, M. and {Aldcroft}, T.~L. and {Cruz}, K.~L. and
         {Robitaille}, T.~P. and {Tollerud}, E.~J. and {Ardelean}, C. and
         {Babej}, T. and {Bach}, Y.~P. and {Bachetti}, M. and {Bakanov}, A.~V. and
         {Bamford}, S.~P. and {Barentsen}, G. and {Barmby}, P. and
         {Baumbach}, A. and {Berry}, K.~L. and {Biscani}, F. and {Boquien}, M. and
         {Bostroem}, K.~A. and {Bouma}, L.~G. and {Brammer}, G.~B. and
         {Bray}, E.~M. and {Breytenbach}, H. and {Buddelmeijer}, H. and
         {Burke}, D.~J. and {Calderone}, G. and {Cano Rodr{\'\i}guez}, J.~L. and
         {Cara}, M. and {Cardoso}, J.~V.~M. and {Cheedella}, S. and {Copin}, Y. and
         {Corrales}, L. and {Crichton}, D. and {D'Avella}, D. and {Deil}, C. and
         {Depagne}, {\'E}. and {Dietrich}, J.~P. and {Donath}, A. and
         {Droettboom}, M. and {Earl}, N. and {Erben}, T. and {Fabbro}, S. and
         {Ferreira}, L.~A. and {Finethy}, T. and {Fox}, R.~T. and
         {Garrison}, L.~H. and {Gibbons}, S.~L.~J. and {Goldstein}, D.~A. and
         {Gommers}, R. and {Greco}, J.~P. and {Greenfield}, P. and
         {Groener}, A.~M. and {Grollier}, F. and {Hagen}, A. and {Hirst}, P. and
         {Homeier}, D. and {Horton}, A.~J. and {Hosseinzadeh}, G. and {Hu}, L. and
         {Hunkeler}, J.~S. and {Ivezi{\'c}}, {\v{Z}}. and {Jain}, A. and
         {Jenness}, T. and {Kanarek}, G. and {Kendrew}, S. and {Kern}, N.~S. and
         {Kerzendorf}, W.~E. and {Khvalko}, A. and {King}, J. and {Kirkby}, D. and
         {Kulkarni}, A.~M. and {Kumar}, A. and {Lee}, A. and {Lenz}, D. and
         {Littlefair}, S.~P. and {Ma}, Z. and {Macleod}, D.~M. and
         {Mastropietro}, M. and {McCully}, C. and {Montagnac}, S. and
         {Morris}, B.~M. and {Mueller}, M. and {Mumford}, S.~J. and {Muna}, D. and
         {Murphy}, N.~A. and {Nelson}, S. and {Nguyen}, G.~H. and
         {Ninan}, J.~P. and {N{\"o}the}, M. and {Ogaz}, S. and {Oh}, S. and
         {Parejko}, J.~K. and {Parley}, N. and {Pascual}, S. and {Patil}, R. and
         {Patil}, A.~A. and {Plunkett}, A.~L. and {Prochaska}, J.~X. and
         {Rastogi}, T. and {Reddy Janga}, V. and {Sabater}, J. and
         {Sakurikar}, P. and {Seifert}, M. and {Sherbert}, L.~E. and
         {Sherwood-Taylor}, H. and {Shih}, A.~Y. and {Sick}, J. and
         {Silbiger}, M.~T. and {Singanamalla}, S. and {Singer}, L.~P. and
         {Sladen}, P.~H. and {Sooley}, K.~A. and {Sornarajah}, S. and
         {Streicher}, O. and {Teuben}, P. and {Thomas}, S.~W. and
         {Tremblay}, G.~R. and {Turner}, J.~E.~H. and {Terr{\'o}n}, V. and
         {van Kerkwijk}, M.~H. and {de la Vega}, A. and {Watkins}, L.~L. and
         {Weaver}, B.~A. and {Whitmore}, J.~B. and {Woillez}, J. and
         {Zabalza}, V. and {Astropy Contributors}},
        title = "{The Astropy Project: Building an Open-science Project and Status of the v2.0 Core Package}",
      journal = {\aj},
         year = 2018,
        month = sep,
       volume = {156},
       number = {3},
          eid = {123},
        pages = {123},
          doi = {10.3847/1538-3881/aabc4f},
archivePrefix = {arXiv},
       eprint = {1801.02634},
 primaryClass = {astro-ph.IM},
       adsurl = {https://ui.adsabs.harvard.edu/abs/2018AJ....156..123A}
}

@ARTICLE{astropy:2022,
       author = {{Astropy Collaboration} and {Price-Whelan}, Adrian M. and {Lim}, Pey Lian and {Earl}, Nicholas and {Starkman}, Nathaniel and {Bradley}, Larry and {Shupe}, David L. and {Patil}, Aarya A. and {Corrales}, Lia and {Brasseur}, C.~E. and {N{\"o}the}, Maximilian and {Donath}, Axel and {Tollerud}, Erik and {Morris}, Brett M. and {Ginsburg}, Adam and {Vaher}, Eero and {Weaver}, Benjamin A. and {Tocknell}, James and {Jamieson}, William and {van Kerkwijk}, Marten H. and {Robitaille}, Thomas P. and {Merry}, Bruce and {Bachetti}, Matteo and {G{\"u}nther}, H. Moritz and {Aldcroft}, Thomas L. and {Alvarado-Montes}, Jaime A. and {Archibald}, Anne M. and {B{\'o}di}, Attila and {Bapat}, Shreyas and {Barentsen}, Geert and {Baz{\'a}n}, Juanjo and {Biswas}, Manish and {Boquien}, M{\'e}d{\'e}ric and {Burke}, D.~J. and {Cara}, Daria and {Cara}, Mihai and {Conroy}, Kyle E. and {Conseil}, Simon and {Craig}, Matthew W. and {Cross}, Robert M. and {Cruz}, Kelle L. and {D'Eugenio}, Francesco and {Dencheva}, Nadia and {Devillepoix}, Hadrien A.~R. and {Dietrich}, J{\"o}rg P. and {Eigenbrot}, Arthur Davis and {Erben}, Thomas and {Ferreira}, Leonardo and {Foreman-Mackey}, Daniel and {Fox}, Ryan and {Freij}, Nabil and {Garg}, Suyog and {Geda}, Robel and {Glattly}, Lauren and {Gondhalekar}, Yash and {Gordon}, Karl D. and {Grant}, David and {Greenfield}, Perry and {Groener}, Austen M. and {Guest}, Steve and {Gurovich}, Sebastian and {Handberg}, Rasmus and {Hart}, Akeem and {Hatfield-Dodds}, Zac and {Homeier}, Derek and {Hosseinzadeh}, Griffin and {Jenness}, Tim and {Jones}, Craig K. and {Joseph}, Prajwel and {Kalmbach}, J. Bryce and {Karamehmetoglu}, Emir and {Ka{\l}uszy{\'n}ski}, Miko{\l}aj and {Kelley}, Michael S.~P. and {Kern}, Nicholas and {Kerzendorf}, Wolfgang E. and {Koch}, Eric W. and {Kulumani}, Shankar and {Lee}, Antony and {Ly}, Chun and {Ma}, Zhiyuan and {MacBride}, Conor and {Maljaars}, Jakob M. and {Muna}, Demitri and {Murphy}, N.~A. and {Norman}, Henrik and {O'Steen}, Richard and {Oman}, Kyle A. and {Pacifici}, Camilla and {Pascual}, Sergio and {Pascual-Granado}, J. and {Patil}, Rohit R. and {Perren}, Gabriel I. and {Pickering}, Timothy E. and {Rastogi}, Tanuj and {Roulston}, Benjamin R. and {Ryan}, Daniel F. and {Rykoff}, Eli S. and {Sabater}, Jose and {Sakurikar}, Parikshit and {Salgado}, Jes{\'u}s and {Sanghi}, Aniket and {Saunders}, Nicholas and {Savchenko}, Volodymyr and {Schwardt}, Ludwig and {Seifert-Eckert}, Michael and {Shih}, Albert Y. and {Jain}, Anany Shrey and {Shukla}, Gyanendra and {Sick}, Jonathan and {Simpson}, Chris and {Singanamalla}, Sudheesh and {Singer}, Leo P. and {Singhal}, Jaladh and {Sinha}, Manodeep and {Sip{\H{o}}cz}, Brigitta M. and {Spitler}, Lee R. and {Stansby}, David and {Streicher}, Ole and {{\v{S}}umak}, Jani and {Swinbank}, John D. and {Taranu}, Dan S. and {Tewary}, Nikita and {Tremblay}, Grant R. and {de Val-Borro}, Miguel and {Van Kooten}, Samuel J. and {Vasovi{\'c}}, Zlatan and {Verma}, Shresth and {de Miranda Cardoso}, Jos{\'e} Vin{\'\i}cius and {Williams}, Peter K.~G. and {Wilson}, Tom J. and {Winkel}, Benjamin and {Wood-Vasey}, W.~M. and {Xue}, Rui and {Yoachim}, Peter and {Zhang}, Chen and {Zonca}, Andrea and {Astropy Project Contributors}},
        title = "{The Astropy Project: Sustaining and Growing a Community-oriented Open-source Project and the Latest Major Release (v5.0) of the Core Package}",
      journal = {\apj},
         year = 2022,
        month = aug,
       volume = {935},
       number = {2},
          eid = {167},
        pages = {167},
          doi = {10.3847/1538-4357/ac7c74},
archivePrefix = {arXiv},
       eprint = {2206.14220},
 primaryClass = {astro-ph.IM},
       adsurl = {https://ui.adsabs.harvard.edu/abs/2022ApJ...935..167A}
}

@Article{matplotlib,
  Author    = {Hunter, J. D.},
  Title     = {Matplotlib: A 2D graphics environment},
  Journal   = {Computing in Science \& Engineering},
  Volume    = {9},
  Number    = {3},
  Pages     = {90--95},
  publisher = {IEEE COMPUTER SOC},
  doi       = {10.1109/MCSE.2007.55},
  year      = 2007
}

@Article{numpy,
 title         = {Array programming with {NumPy}},
 author        = {Charles R. Harris and K. Jarrod Millman and St{\'{e}}fan J.
                 van der Walt and Ralf Gommers and Pauli Virtanen and David
                 Cournapeau and Eric Wieser and Julian Taylor and Sebastian
                 Berg and Nathaniel J. Smith and Robert Kern and Matti Picus
                 and Stephan Hoyer and Marten H. van Kerkwijk and Matthew
                 Brett and Allan Haldane and Jaime Fern{\'{a}}ndez del
                 R{\'{i}}o and Mark Wiebe and Pearu Peterson and Pierre
                 G{\'{e}}rard-Marchant and Kevin Sheppard and Tyler Reddy and
                 Warren Weckesser and Hameer Abbasi and Christoph Gohlke and
                 Travis E. Oliphant},
 year          = {2020},
 month         = sep,
 journal       = {Nature},
 volume        = {585},
 number        = {7825},
 pages         = {357--362},
 doi           = {10.1038/s41586-020-2649-2},
 publisher     = {Springer Science and Business Media {LLC}},
 url           = {https://doi.org/10.1038/s41586-020-2649-2}
}

@article{Winters_2019,
   title={The Solar Neighborhood. XLV. The Stellar Multiplicity Rate of M Dwarfs Within 25 pc},
   volume={157},
   ISSN={1538-3881},
   url={http://dx.doi.org/10.3847/1538-3881/ab05dc},
   DOI={10.3847/1538-3881/ab05dc},
   number={6},
   journal={The Astronomical Journal},
   publisher={American Astronomical Society},
   author={Winters, Jennifer G. and Henry, Todd J. and Jao, Wei-Chun and Subasavage, John P. and Chatelain, Joseph P. and Slatten, Ken and Riedel, Adric R. and Silverstein, Michele L. and Payne, Matthew J.},
   year={2019},
   month=may, pages={216} }

@ARTICLE{Chabrier_2003,
       author = {{Chabrier}, Gilles},
        title = "{Galactic Stellar and Substellar Initial Mass Function}",
      journal = {\pasp},
         year = 2003,
        month = jul,
       volume = {115},
       number = {809},
        pages = {763-795},
          doi = {10.1086/376392},
archivePrefix = {arXiv},
       eprint = {astro-ph/0304382},
 primaryClass = {astro-ph},
       adsurl = {https://ui.adsabs.harvard.edu/abs/2003PASP..115..763C}
}

@ARTICLE{Carr_2021,
       author = {{Carr}, Bernard and {Kohri}, Kazunori and {Sendouda}, Yuuiti and {Yokoyama}, Jun'ichi},
        title = "{Constraints on primordial black holes}",
      journal = {Reports on Progress in Physics},
         year = 2021,
        month = nov,
       volume = {84},
       number = {11},
          eid = {116902},
        pages = {116902},
          doi = {10.1088/1361-6633/ac1e31},
archivePrefix = {arXiv},
       eprint = {2002.12778},
 primaryClass = {astro-ph.CO},
       adsurl = {https://ui.adsabs.harvard.edu/abs/2021RPPh...84k6902C}
}

@article{Mroz_2024,
   title={Limits on Planetary-mass Primordial Black Holes from the OGLE High-cadence Survey of the Magellanic Clouds},
   volume={976},
   ISSN={2041-8213},
   url={http://dx.doi.org/10.3847/2041-8213/ad8e68},
   DOI={10.3847/2041-8213/ad8e68},
   number={1},
   journal={The Astrophysical Journal Letters},
   publisher={American Astronomical Society},
   author={Mróz, Przemek and Udalski, Andrzej and Szymański, Michał K. and Soszyński, Igor and Pietrukowicz, Paweł and Kozłowski, Szymon and Poleski, Radosław and Skowron, Jan and Ulaczyk, Krzysztof and Gromadzki, Mariusz and Rybicki, Krzysztof and Iwanek, Patryk and Wrona, Marcin and Mróz, Mateusz J.},
   year={2024},
   month=nov, pages={L19} }

@article{Montero_Camacho_2019,
   title={Revisiting constraints on asteroid-mass primordial black holes as dark matter candidates},
   volume={2019},
   ISSN={1475-7516},
   url={http://dx.doi.org/10.1088/1475-7516/2019/08/031},
   DOI={10.1088/1475-7516/2019/08/031},
   number={08},
   journal={Journal of Cosmology and Astroparticle Physics},
   publisher={IOP Publishing},
   author={Montero-Camacho, Paulo and Fang, Xiao and Vasquez, Gabriel and Silva, Makana and Hirata, Christopher M.},
   year={2019},
   month=aug, pages={031–031} }

@ARTICLE{Gannon_2026,
       author = {{Gannon}, Jonah S. and {Ferr{\'e}-Mateu}, Anna and {Forbes}, Duncan A.},
        title = "{The Dawes Review 14: A Decade of Ultra-Diffuse Galaxies}",
      journal = {arXiv e-prints},
         year = 2026,
        month = feb,
          eid = {arXiv:2602.21875},
        pages = {arXiv:2602.21875},
          doi = {10.48550/arXiv.2602.21875},
archivePrefix = {arXiv},
       eprint = {2602.21875},
 primaryClass = {astro-ph.GA},
       adsurl = {https://ui.adsabs.harvard.edu/abs/2026arXiv260221875G}
}

@ARTICLE{Chang_Refsdal_1979,
       author = {{Chang}, K. and {Refsdal}, S.},
        title = "{Flux variations of QSO 0957 + 561 A, B and image splitting by stars near the light path}",
      journal = {\nat},
         year = 1979,
        month = dec,
       volume = {282},
       number = {5739},
        pages = {561-564},
          doi = {10.1038/282561a0},
       adsurl = {https://ui.adsabs.harvard.edu/abs/1979Natur.282..561C}
}

@ARTICLE{Sersic_1963,
       author = {{S{\'e}rsic}, J.~L.},
        title = "{Influence of the atmospheric and instrumental dispersion on the brightness distribution in a galaxy}",
      journal = {Boletin de la Asociacion Argentina de Astronomia La Plata Argentina},
         year = 1963,
        month = feb,
       volume = {6},
        pages = {41-43},
       adsurl = {https://ui.adsabs.harvard.edu/abs/1963BAAA....6...41S}
}

@article{Roman_2017,
    author = {Roman, Javier and Trujillo, Ignacio},
    title = {Ultra-diffuse galaxies outside clusters: clues to their formation and evolution},
    journal = {Monthly Notices of the Royal Astronomical Society},
    volume = {468},
    number = {4},
    pages = {4039-4047},
    year = {2017},
    month = {07},
    issn = {0035-8711},
    doi = {10.1093/mnras/stx694},
    url = {https://doi.org/10.1093/mnras/stx694},
    eprint = {https://academic.oup.com/mnras/article-pdf/468/4/4039/13953618/stx694.pdf},
}

@article{Koda_2015,
doi = {10.1088/2041-8205/807/1/L2},
url = {https://doi.org/10.1088/2041-8205/807/1/L2},
year = {2015},
month = {jun},
publisher = {The American Astronomical Society},
volume = {807},
number = {1},
pages = {L2},
author = {Koda, Jin and Yagi, Masafumi and Yamanoi, Hitomi and Komiyama, Yutaka},
title = {APPROXIMATELY A THOUSAND ULTRA-DIFFUSE GALAXIES IN THE COMA CLUSTER},
journal = {The Astrophysical Journal Letters}
}

@article{van_Dokkum_2018_distance,
doi = {10.3847/2041-8213/aada4d},
url = {https://doi.org/10.3847/2041-8213/aada4d},
year = {2018},
month = {aug},
publisher = {The American Astronomical Society},
volume = {864},
number = {1},
pages = {L18},
author = {van Dokkum, Pieter and Danieli, Shany and Cohen, Yotam and Romanowsky, Aaron J. and Conroy, Charlie},
title = {The Distance of the Dark Matter Deficient Galaxy NGC 1052–DF2},
journal = {The Astrophysical Journal Letters}
}

@ARTICLE{Beasley_2025,
       author = {{Beasley}, M.~A. and {Fahrion}, K. and {Guerra Arencibia}, S. and {Gvozdenko}, A. and {Montes}, M.},
        title = "{A new way to measure the distance to NGC1052-DF2}",
      journal = {\aap},
         year = 2025,
        month = may,
       volume = {697},
          eid = {A144},
        pages = {A144},
          doi = {10.1051/0004-6361/202452446},
archivePrefix = {arXiv},
       eprint = {2503.03403},
 primaryClass = {astro-ph.GA},
       adsurl = {https://ui.adsabs.harvard.edu/abs/2025A&A...697A.144B}
}

@ARTICLE{Fahrion_2025,
       author = {{Fahrion}, K. and {Beasley}, M.~A. and {Gvozdenko}, A. and {Guerra Arencibia}, S. and {Jerabkova}, T. and {Fensch}, J. and {Emsellem}, E.},
        title = "{Revisiting the globular clusters of NGC 1052-DF2}",
      journal = {\aap},
         year = 2025,
        month = may,
       volume = {697},
          eid = {A145},
        pages = {A145},
          doi = {10.1051/0004-6361/202452454},
archivePrefix = {arXiv},
       eprint = {2503.03404},
 primaryClass = {astro-ph.GA},
       adsurl = {https://ui.adsabs.harvard.edu/abs/2025A&A...697A.145F}
}

@ARTICLE{Raghavan_2010,
       author = {{Raghavan}, Deepak and {McAlister}, Harold A. and {Henry}, Todd J. and {Latham}, David W. and {Marcy}, Geoffrey W. and {Mason}, Brian D. and {Gies}, Douglas R. and {White}, Russel J. and {ten Brummelaar}, Theo A.},
        title = "{A Survey of Stellar Families: Multiplicity of Solar-type Stars}",
      journal = {\apjs},
         year = 2010,
        month = sep,
       volume = {190},
       number = {1},
        pages = {1-42},
          doi = {10.1088/0067-0049/190/1/1},
archivePrefix = {arXiv},
       eprint = {1007.0414},
 primaryClass = {astro-ph.SR},
       adsurl = {https://ui.adsabs.harvard.edu/abs/2010ApJS..190....1R}
}

@ARTICLE{Diego_2024_BF,
       author = {{Diego}, Jose M. and {Li}, Sung Kei and {Meena}, Ashish K. and {Niemiec}, Anna and {Acebron}, Ana and {Jauzac}, Mathilde and {Struble}, Mitchell F. and {Amruth}, Alfred and {Broadhurst}, Tom J. and {Cerny}, Catherine and {Ebeling}, Harald and {Filippenko}, Alexei V. and {Jullo}, Eric and {Kelly}, Patrick and {Koekemoer}, Anton M. and {Lagattuta}, David and {Lim}, Jeremy and {Limousin}, Marceau and {Mahler}, Guillaume and {Patel}, Nency and {Remolina}, Juan and {Richard}, Johan and {Sharon}, Keren and {Steinhardt}, Charles and {Umetsu}, Keiichi and {Williams}, Liliya and {Zitrin}, Adi and {Palencia}, Jose Mar{\'\i}a and {Dai}, Liang and {Ji}, Lingyuan and {Pascale}, Massimo},
        title = "{BUFFALO/Flashlights: Constraints on the abundance of lensed supergiant stars in the Spock galaxy at redshift 1}",
      journal = {\aap},
         year = 2024,
        month = jan,
       volume = {681},
          eid = {A124},
        pages = {A124},
          doi = {10.1051/0004-6361/202346761},
archivePrefix = {arXiv},
       eprint = {2304.09222},
 primaryClass = {astro-ph.GA},
       adsurl = {https://ui.adsabs.harvard.edu/abs/2024A&A...681A.124D}
}

@ARTICLE{Li_2025_IMF,
       author = {{Li}, Sung Kei and {Diego}, Jose M. and {Meena}, Ashish K. and {Lim}, Jeremy and {Fung}, Leo W.~H. and {Levitskiy}, Arsen and {Nianias}, James and {Palencia}, Jose M. and {Williams}, Hayley and {Zhang}, Jiashuo and {Amruth}, Alfred and {Broadhurst}, Thomas J. and {Chen}, WenLei and {Filippenko}, Alexei V. and {Kelly}, Patrick L. and {Koekemoer}, Anton M. and {Perera}, Derek and {Sun}, Bangzheng and {Williams}, Liliya L.~R. and {Windhorst}, Rogier A. and {Yan}, Haojin and {Zitrin}, Adi},
        title = "{Constraining the z {\ensuremath{\sim}} 1 Initial Mass Function with HST and JWST Lensed Stars in MACS J0416.1{\ensuremath{-}}2403}",
      journal = {\apj},
         year = 2025,
        month = aug,
       volume = {988},
       number = {2},
          eid = {178},
        pages = {178},
          doi = {10.3847/1538-4357/ade4bd},
archivePrefix = {arXiv},
       eprint = {2504.06992},
 primaryClass = {astro-ph.CO},
       adsurl = {https://ui.adsabs.harvard.edu/abs/2025ApJ...988..178L}
}

@ARTICLE{Diego_2024_3M,
       author = {{Diego}, Jose M. and {Li}, Sung Kei and {Amruth}, Alfred and {Meena}, Ashish K. and {Broadhurst}, Tom J. and {Kelly}, Patrick L. and {Filippenko}, Alexei V. and {Williams}, Liliya L.~R. and {Zitrin}, Adi and {Harris}, William E. and {Reina-Campos}, Marta and {Giocoli}, Carlo and {Dai}, Liang and {Struble}, Mitchell F. and {Treu}, Tommaso and {Fudamoto}, Yoshinobu and {Gilman}, Daniel and {Koekemoer}, Anton M. and {Lim}, Jeremy and {Palencia}, Jose Mar{\'\i}a and {Sun}, Fengwu and {Windhorst}, Rogier A.},
        title = "{Imaging dark matter at the smallest scales with z {\ensuremath{\approx}} 1 lensed stars}",
      journal = {\aap},
         year = 2024,
        month = sep,
       volume = {689},
          eid = {A167},
        pages = {A167},
          doi = {10.1051/0004-6361/202450474},
archivePrefix = {arXiv},
       eprint = {2404.08033},
 primaryClass = {astro-ph.CO},
       adsurl = {https://ui.adsabs.harvard.edu/abs/2024A&A...689A.167D}
}

@ARTICLE{Broadhurst_2025,
       author = {{Broadhurst}, Tom and {Li}, Sung Kei and {Alfred}, Amruth and {Diego}, Jose M. and {Morilla}, Paloma and {Kelly}, Patrick L. and {Sun}, Fengwu and {Oguri}, Masamune and {Williams}, Hayley and {Windhorst}, Rogier and {Zitrin}, Adi and {Abe}, Katsuya T. and {Chen}, Wenlei and {Dai}, Liang and {Fudamoto}, Yoshinobu and {Kawai}, Hiroki and {Lim}, Jeremy and {Liu}, Tao and {Meena}, Ashish K. and {Palencia}, Jose M. and {Smoot}, George F. and {Williams}, Liliya L.~R.},
        title = "{Dark Matter Distinguished by Skewed Microlensing in the ``Dragon Arc''}",
      journal = {\apjl},
         year = 2025,
        month = jan,
       volume = {978},
       number = {1},
          eid = {L5},
        pages = {L5},
          doi = {10.3847/2041-8213/ad9aa8},
archivePrefix = {arXiv},
       eprint = {2405.19422},
 primaryClass = {astro-ph.CO},
       adsurl = {https://ui.adsabs.harvard.edu/abs/2025ApJ...978L...5B}
}

@ARTICLE{Gaudi_2012,
       author = {{Gaudi}, B. Scott},
        title = "{Microlensing Surveys for Exoplanets}",
      journal = {\araa},
         year = 2012,
        month = sep,
       volume = {50},
        pages = {411-453},
          doi = {10.1146/annurev-astro-081811-125518},
       adsurl = {https://ui.adsabs.harvard.edu/abs/2012ARA&A..50..411G}
}

@article{Calcino_2018,
    author = {Calcino, Josh and Garcia-Bellido, Juan and Davis, Tamara M},
    title = {Updating the MACHO fraction of the Milky Way dark halowith improved mass models},
    journal = {Monthly Notices of the Royal Astronomical Society},
    volume = {479},
    number = {3},
    pages = {2889-2905},
    year = {2018},
    month = {05},
    issn = {0035-8711},
    doi = {10.1093/mnras/sty1368},
    url = {https://doi.org/10.1093/mnras/sty1368},
    eprint = {https://academic.oup.com/mnras/article-pdf/479/3/2889/25149543/sty1368.pdf},
}

@ARTICLE{Brocklehurst_1971,
       author = {{Brocklehurst}, M.},
        title = "{Calculations of level populations for the low levels of hydrogenic ions in gaseous nebulae.}",
      journal = {\mnras},
         year = 1971,
        month = jan,
       volume = {153},
        pages = {471},
          doi = {10.1093/mnras/153.4.471},
       adsurl = {https://ui.adsabs.harvard.edu/abs/1971MNRAS.153..471B}
}

@ARTICLE{Zeimann_2014,
       author = {{Zeimann}, Gregory R. and {Ciardullo}, Robin and {Gebhardt}, Henry and {Gronwall}, Caryl and {Schneider}, Donald P. and {Hagen}, Alex and {Bridge}, Joanna S. and {Feldmeier}, John and {Trump}, Jonathan R.},
        title = "{3D-HST Emission Line Galaxies at z \raisebox{-0.5ex}\textasciitilde 2: Discrepancies in the Optical/UV Star Formation Rates}",
      journal = {\apj},
         year = 2014,
        month = aug,
       volume = {790},
       number = {2},
          eid = {113},
        pages = {113},
          doi = {10.1088/0004-637X/790/2/113},
archivePrefix = {arXiv},
       eprint = {1406.3355},
 primaryClass = {astro-ph.GA},
       adsurl = {https://ui.adsabs.harvard.edu/abs/2014ApJ...790..113Z}
}

@ARTICLE{Fensch_2019,
       author = {{Fensch}, J{\'e}r{\'e}my and {van der Burg}, Remco F.~J. and {Je{\v{r}}{\'a}bkov{\'a}}, Tereza and {Emsellem}, Eric and {Zanella}, Anita and {Agnello}, Adriano and {Hilker}, Michael and {M{\"u}ller}, Oliver and {Rejkuba}, Marina and {Duc}, Pierre-Alain and {Durrell}, Patrick and {Habas}, Rebecca and {Lim}, Sungsoon and {Marleau}, Francine R. and {Peng}, Eric W. and {S{\'a}nchez Janssen}, Rub{\'e}n},
        title = "{The ultra-diffuse galaxy NGC 1052-DF2 with MUSE. II. The population of DF2: stars, clusters, and planetary nebulae}",
      journal = {\aap},
         year = 2019,
        month = may,
       volume = {625},
          eid = {A77},
        pages = {A77},
          doi = {10.1051/0004-6361/201834911},
archivePrefix = {arXiv},
       eprint = {1812.07346},
 primaryClass = {astro-ph.GA},
       adsurl = {https://ui.adsabs.harvard.edu/abs/2019A&A...625A..77F}
}

@ARTICLE{Kelly_2022_Flashlights,
       author = {{Kelly}, Patrick L. and {Chen}, Wenlei and {Alfred}, Amruth and {Broadhurst}, Thomas J. and {Diego}, Jose M. and {Emami}, Najmeh and {Filippenko}, Alexei V. and {Keen}, Allison and {Li}, Sung Kei and {Lim}, Jeremy and {Meena}, Ashish K. and {Oguri}, Masamune and {Scarlata}, Claudia and {Treu}, Tommaso and {Williams}, Hayley and {Williams}, Liliya L.~R. and {Zhou}, Rui and {Zitrin}, Adi and {Foley}, Ryan J. and {Jha}, Saurabh W. and {Kaiser}, Nick and {Mehta}, Vihang and {Rieck}, Steven and {Salo}, Laura and {Smith}, Nathan and {Weisz}, Daniel R.},
        title = "{Flashlights: More than A Dozen High-Significance Microlensing Events of Extremely Magnified Stars in Galaxies at Redshifts z=0.7-1.5}",
      journal = {arXiv e-prints},
         year = 2022,
        month = nov,
          eid = {arXiv:2211.02670},
        pages = {arXiv:2211.02670},
          doi = {10.48550/arXiv.2211.02670},
archivePrefix = {arXiv},
       eprint = {2211.02670},
 primaryClass = {astro-ph.CO},
       adsurl = {https://ui.adsabs.harvard.edu/abs/2022arXiv221102670K}
}

@ARTICLE{Li_2025_Horseshoe,
       author = {{Li}, Sung Kei and {Weisenbach}, Luke and {Collett}, Thomas E. and {Diego}, Jose M. and {Lim}, Jeremy and {Broadhurst}, Thomas J. and {Chow}, Alex and {Enzi}, Wolfgang J.~R. and {Kelly}, Patrick L. and {Melo-Carneiro}, Carlos R. and {Palencia}, Jose M. and {Williams}, Liliya L.~R. and {Zhang}, Jiashuo},
        title = "{Lensed stars in galaxy─galaxy strong lensing ─ a JWST prediction for the Cosmic Horseshoe}",
      journal = {\mnras},
         year = 2025,
        month = dec,
       volume = {544},
       number = {4},
        pages = {4469-4481},
          doi = {10.1093/mnras/staf1959},
archivePrefix = {arXiv},
       eprint = {2509.16154},
 primaryClass = {astro-ph.CO},
       adsurl = {https://ui.adsabs.harvard.edu/abs/2025MNRAS.544.4469L}
}

@ARTICLE{Chamba_2020,
       author = {{Chamba}, Nushkia and {Trujillo}, Ignacio and {Knapen}, Johan H.},
        title = "{Are ultra-diffuse galaxies Milky Way-sized?}",
      journal = {\aap},
         year = 2020,
        month = jan,
       volume = {633},
          eid = {L3},
        pages = {L3},
          doi = {10.1051/0004-6361/201936821},
archivePrefix = {arXiv},
       eprint = {2001.02691},
 primaryClass = {astro-ph.GA},
       adsurl = {https://ui.adsabs.harvard.edu/abs/2020A&A...633L...3C}
}

@ARTICLE{Benavides_2021,
       author = {{Benavides}, Jos{\'e} A. and {Sales}, Laura V. and {Abadi}, Mario. G. and {Pillepich}, Annalisa and {Nelson}, Dylan and {Marinacci}, Federico and {Cooper}, Michael and {Pakmor}, Ruediger and {Torrey}, Paul and {Vogelsberger}, Mark and {Hernquist}, Lars},
        title = "{Quiescent ultra-diffuse galaxies in the field originating from backsplash orbits}",
      journal = {Nature Astronomy},
         year = 2021,
        month = sep,
       volume = {5},
        pages = {1255-1260},
          doi = {10.1038/s41550-021-01458-1},
archivePrefix = {arXiv},
       eprint = {2109.01677},
 primaryClass = {astro-ph.GA},
       adsurl = {https://ui.adsabs.harvard.edu/abs/2021NatAs...5.1255B}
}

@ARTICLE{Planck18,
       author = {{Planck Collaboration} and {Aghanim}, N. and {Akrami}, Y. and {Ashdown}, M. and {Aumont}, J. and {Baccigalupi}, C. and {Ballardini}, M. and {Banday}, A.~J. and {Barreiro}, R.~B. and {Bartolo}, N. and {Basak}, S. and {Battye}, R. and {Benabed}, K. and {Bernard}, J.-P. and {Bersanelli}, M. and {Bielewicz}, P. and {Bock}, J.~J. and {Bond}, J.~R. and {Borrill}, J. and {Bouchet}, F.~R. and {Boulanger}, F. and {Bucher}, M. and {Burigana}, C. and {Butler}, R.~C. and {Calabrese}, E. and {Cardoso}, J.-F. and {Carron}, J. and {Challinor}, A. and {Chiang}, H.~C. and {Chluba}, J. and {Colombo}, L.~P.~L. and {Combet}, C. and {Contreras}, D. and {Crill}, B.~P. and {Cuttaia}, F. and {de Bernardis}, P. and {de Zotti}, G. and {Delabrouille}, J. and {Delouis}, J.-M. and {Di Valentino}, E. and {Diego}, J.~M. and {Dor{\'e}}, O. and {Douspis}, M. and {Ducout}, A. and {Dupac}, X. and {Dusini}, S. and {Efstathiou}, G. and {Elsner}, F. and {En{\ss}lin}, T.~A. and {Eriksen}, H.~K. and {Fantaye}, Y. and {Farhang}, M. and {Fergusson}, J. and {Fernandez-Cobos}, R. and {Finelli}, F. and {Forastieri}, F. and {Frailis}, M. and {Fraisse}, A.~A. and {Franceschi}, E. and {Frolov}, A. and {Galeotta}, S. and {Galli}, S. and {Ganga}, K. and {G{\'e}nova-Santos}, R.~T. and {Gerbino}, M. and {Ghosh}, T. and {Gonz{\'a}lez-Nuevo}, J. and {G{\'o}rski}, K.~M. and {Gratton}, S. and {Gruppuso}, A. and {Gudmundsson}, J.~E. and {Hamann}, J. and {Handley}, W. and {Hansen}, F.~K. and {Herranz}, D. and {Hildebrandt}, S.~R. and {Hivon}, E. and {Huang}, Z. and {Jaffe}, A.~H. and {Jones}, W.~C. and {Karakci}, A. and {Keih{\"a}nen}, E. and {Keskitalo}, R. and {Kiiveri}, K. and {Kim}, J. and {Kisner}, T.~S. and {Knox}, L. and {Krachmalnicoff}, N. and {Kunz}, M. and {Kurki-Suonio}, H. and {Lagache}, G. and {Lamarre}, J.-M. and {Lasenby}, A. and {Lattanzi}, M. and {Lawrence}, C.~R. and {Le Jeune}, M. and {Lemos}, P. and {Lesgourgues}, J. and {Levrier}, F. and {Lewis}, A. and {Liguori}, M. and {Lilje}, P.~B. and {Lilley}, M. and {Lindholm}, V. and {L{\'o}pez-Caniego}, M. and {Lubin}, P.~M. and {Ma}, Y.-Z. and {Mac{\'\i}as-P{\'e}rez}, J.~F. and {Maggio}, G. and {Maino}, D. and {Mandolesi}, N. and {Mangilli}, A. and {Marcos-Caballero}, A. and {Maris}, M. and {Martin}, P.~G. and {Martinelli}, M. and {Mart{\'\i}nez-Gonz{\'a}lez}, E. and {Matarrese}, S. and {Mauri}, N. and {McEwen}, J.~D. and {Meinhold}, P.~R. and {Melchiorri}, A. and {Mennella}, A. and {Migliaccio}, M. and {Millea}, M. and {Mitra}, S. and {Miville-Desch{\^e}nes}, M.-A. and {Molinari}, D. and {Montier}, L. and {Morgante}, G. and {Moss}, A. and {Natoli}, P. and {N{\o}rgaard-Nielsen}, H.~U. and {Pagano}, L. and {Paoletti}, D. and {Partridge}, B. and {Patanchon}, G. and {Peiris}, H.~V. and {Perrotta}, F. and {Pettorino}, V. and {Piacentini}, F. and {Polastri}, L. and {Polenta}, G. and {Puget}, J.-L. and {Rachen}, J.~P. and {Reinecke}, M. and {Remazeilles}, M. and {Renzi}, A. and {Rocha}, G. and {Rosset}, C. and {Roudier}, G. and {Rubi{\~n}o-Mart{\'\i}n}, J.~A. and {Ruiz-Granados}, B. and {Salvati}, L. and {Sandri}, M. and {Savelainen}, M. and {Scott}, D. and {Shellard}, E.~P.~S. and {Sirignano}, C. and {Sirri}, G. and {Spencer}, L.~D. and {Sunyaev}, R. and {Suur-Uski}, A.-S. and {Tauber}, J.~A. and {Tavagnacco}, D. and {Tenti}, M. and {Toffolatti}, L. and {Tomasi}, M. and {Trombetti}, T. and {Valenziano}, L. and {Valiviita}, J. and {Van Tent}, B. and {Vibert}, L. and {Vielva}, P. and {Villa}, F. and {Vittorio}, N. and {Wandelt}, B.~D. and {Wehus}, I.~K. and {White}, M. and {White}, S.~D.~M. and {Zacchei}, A. and {Zonca}, A.},
        title = "{Planck 2018 results. VI. Cosmological parameters}",
      journal = {\aap},
         year = 2020,
        month = sep,
       volume = {641},
          eid = {A6},
        pages = {A6},
          doi = {10.1051/0004-6361/201833910},
archivePrefix = {arXiv},
       eprint = {1807.06209},
 primaryClass = {astro-ph.CO},
       adsurl = {https://ui.adsabs.harvard.edu/abs/2020A&A...641A...6P}
}

@ARTICLE{Oke_1983,
       author = {{Oke}, J.~B. and {Gunn}, J.~E.},
        title = "{Secondary standard stars for absolute spectrophotometry.}",
      journal = {\apj},
         year = 1983,
        month = mar,
       volume = {266},
        pages = {713-717},
          doi = {10.1086/160817},
       adsurl = {https://ui.adsabs.harvard.edu/abs/1983ApJ...266..713O}
}

@ARTICLE{Van_Dokkum_2015,
       author = {{van Dokkum}, Pieter G. and {Abraham}, Roberto and {Merritt}, Allison and {Zhang}, Jielai and {Geha}, Marla and {Conroy}, Charlie},
        title = "{Forty-seven Milky Way-sized, Extremely Diffuse Galaxies in the Coma Cluster}",
      journal = {\apjl},
         year = 2015,
        month = jan,
       volume = {798},
       number = {2},
          eid = {L45},
        pages = {L45},
          doi = {10.1088/2041-8205/798/2/L45},
archivePrefix = {arXiv},
       eprint = {1410.8141},
 primaryClass = {astro-ph.GA},
       adsurl = {https://ui.adsabs.harvard.edu/abs/2015ApJ...798L..45V}
}

@ARTICLE{Sandage_1984,
       author = {{Sandage}, A. and {Binggeli}, B.},
        title = "{Studies of the Virgo cluster. III. A classification system and an illustrated Atlas of Virgo cluster dwarf galaxies.}",
      journal = {\aj},
         year = 1984,
        month = jul,
       volume = {89},
        pages = {919-931},
          doi = {10.1086/113588},
       adsurl = {https://ui.adsabs.harvard.edu/abs/1984AJ.....89..919S}
}

@ARTICLE{Yan_2023,
       author = {{Yan}, Haojing and {Ma}, Zhiyuan and {Sun}, Bangzheng and {Wang}, Lifan and {Kelly}, Patrick and {Diego}, Jos{\'e} M. and {Cohen}, Seth H. and {Windhorst}, Rogier A. and {Jansen}, Rolf A. and {Grogin}, Norman A. and {Beacom}, John F. and {Conselice}, Christopher J. and {Driver}, Simon P. and {Frye}, Brenda and {Coe}, Dan and {Marshall}, Madeline A. and {Koekemoer}, Anton and {Willmer}, Christopher N.~A. and {Robotham}, Aaron and {D'Silva}, Jordan C.~J. and {Summers}, Jake and {Nonino}, Mario and {Pirzkal}, Nor and {Ryan}, Russell E. and {Ortiz}, Rafael and {Tompkins}, Scott and {Bhatawdekar}, Rachana A. and {Cheng}, Cheng and {Zitrin}, Adi and {Willner}, S.~P.},
        title = "{JWST's PEARLS: Transients in the MACS J0416.1-2403 Field}",
      journal = {\apjs},
         year = 2023,
        month = dec,
       volume = {269},
       number = {2},
          eid = {43},
        pages = {43},
          doi = {10.3847/1538-4365/ad0298},
archivePrefix = {arXiv},
       eprint = {2307.07579},
 primaryClass = {astro-ph.GA},
       adsurl = {https://ui.adsabs.harvard.edu/abs/2023ApJS..269...43Y}
}

@ARTICLE{Fudamoto_2025,
       author = {{Fudamoto}, Yoshinobu and {Sun}, Fengwu and {Diego}, Jose M. and {Dai}, Liang and {Oguri}, Masamune and {Zitrin}, Adi and {Zackrisson}, Erik and {Jauzac}, Mathilde and {Lagattuta}, David J. and {Egami}, Eiichi and {Iani}, Edoardo and {Windhorst}, Rogier A. and {Abe}, Katsuya T. and {Bauer}, Franz Erik and {Bian}, Fuyan and {Bhatawdekar}, Rachana and {Broadhurst}, Thomas J. and {Cai}, Zheng and {Chen}, Chian-Chou and {Chen}, Wenlei and {Cohen}, Seth H. and {Conselice}, Christopher J. and {Espada}, Daniel and {Foo}, Nicholas and {Frye}, Brenda L. and {Fujimoto}, Seiji and {Furtak}, Lukas J. and {Golubchik}, Miriam and {Hsiao}, Tiger Yu-Yang and {Jolly}, Jean-Baptiste and {Kawai}, Hiroki and {Kelly}, Patrick L. and {Koekemoer}, Anton M. and {Kohno}, Kotaro and {Kokorev}, Vasily and {Li}, Mingyu and {Li}, Zihao and {Lin}, Xiaojing and {Magdis}, Georgios E. and {Meena}, Ashish K. and {Niemiec}, Anna and {Nabizadeh}, Armin and {Richard}, Johan and {Steinhardt}, Charles L. and {Wu}, Yunjing and {Zhu}, Yongda and {Zou}, Siwei},
        title = "{Identification of more than 40 gravitationally magnified stars in a galaxy at redshift 0.725}",
      journal = {Nature Astronomy},
         year = 2025,
        month = mar,
       volume = {9},
        pages = {428-437},
          doi = {10.1038/s41550-024-02432-3},
archivePrefix = {arXiv},
       eprint = {2404.08045},
 primaryClass = {astro-ph.GA},
       adsurl = {https://ui.adsabs.harvard.edu/abs/2025NatAs...9..428F}
}

@ARTICLE{Udalski_1994,
       author = {{Udalski}, A. and {Szymanski}, M. and {Kaluzny}, J. and {Kubiak}, M. and {Mateo}, M. and {Krzeminski}, W. and {Paczynski}, B.},
        title = "{The Optical Gravitational Lensing Experiment. The Early Warning System: Real Time Microlensing}",
      journal = {\actaa},
         year = 1994,
        month = jul,
       volume = {44},
        pages = {227-234},
          doi = {10.48550/arXiv.astro-ph/9408026},
archivePrefix = {arXiv},
       eprint = {astro-ph/9408026},
 primaryClass = {astro-ph},
       adsurl = {https://ui.adsabs.harvard.edu/abs/1994AcA....44..227U}
}

@ARTICLE{Novati_2008,
       author = {{Calchi Novati}, S. and {de Luca}, F. and {Jetzer}, Ph. and {Mancini}, L. and {Scarpetta}, G.},
        title = "{Microlensing constraints on the Galactic bulge initial mass function}",
      journal = {\aap},
         year = 2008,
        month = mar,
       volume = {480},
       number = {3},
        pages = {723-733},
          doi = {10.1051/0004-6361:20078439},
archivePrefix = {arXiv},
       eprint = {0711.3758},
 primaryClass = {astro-ph},
       adsurl = {https://ui.adsabs.harvard.edu/abs/2008A&A...480..723C}
}

@misc{marleau2025euclidquickdatarelease,
      title={Euclid: Quick Data Release (Q1) -- A census of dwarf galaxies across a range of distances and environments}, 
      author={F. R. Marleau and R. Habas and D. Carollo and C. Tortora and P. -A. Duc and E. Sola and T. Saifollahi and M. Fügenschuh and M. Walmsley and R. Zöller and A. Ferré-Mateu and M. Cantiello and M. Urbano and E. Saremi and R. Ragusa and R. Laureijs and M. Hilker and O. Müller and M. Poulain and R. F. Peletier and S. J. Sprenger and O. Marchal and N. Aghanim and B. Altieri and A. Amara and S. Andreon and N. Auricchio and H. Aussel and C. Baccigalupi and M. Baldi and A. Balestra and S. Bardelli and A. Basset and P. Battaglia and R. Bender and A. Biviano and A. Bonchi and D. Bonino and E. Branchini and M. Brescia and J. Brinchmann and S. Camera and G. Cañas-Herrera and V. Capobianco and C. Carbone and J. Carretero and S. Casas and M. Castellano and G. Castignani and S. Cavuoti and K. C. Chambers and A. Cimatti and C. Colodro-Conde and G. Congedo and C. J. Conselice L. Conversi and Y. Copin and L. Corcione and F. Courbin and H. M. Courtois and M. Cropper and J. -C. Cuillandre and A. Da Silva and H. Degaudenzi and G. De Lucia and A. M. Di Giorgio and C. Dolding and H. Dole and F. Dubath and X. Dupac and S. Dusini and S. Escoffier and M. Fabricius and M. Farina and F. Faustini and S. Ferriol and P. Fosalba and S. Fotopoulou and M. Frailis and E. Franceschi and P. Franzetti and M. Fumana and S. Galeotta and K. George and B. Gillis and C. Giocoli and B. R. Granett and A. Grazian and F. Grupp and S. Gwyn and S. V. H. Haugan and J. Hoar and H. Hoekstra and W. Holmes and F. Hormuth and A. Hornstrup and P. Hudelot and K. Jahnke and M. Jhabvala and B. Joachimi and E. Keihänen and S. Kermiche and A. Kiessling and B. Kubik and M. Kümmel and M. Kunz and H. Kurki-Suonio and O. Lahav and Q. Le Boulc'h and A. M. C. Le Brun and D. Le Mignant and S. Ligori and P. B. Lilje and V. Lindholm and I. Lloro and G. Mainetti and D. Maino and E. Maiorano and O. Mansutti and S. Marcin and O. Marggraf and M. Martinelli and N. Martinet and F. Marulli and R. Massey and S. Maurogordato and H. J. McCracken and E. Medinaceli and S. Mei and M. Melchior and Y. Mellier and M. Meneghetti and E. Merlin and G. Meylan and A. Mora and M. Moresco and L. Moscardini and R. Nakajima and C. Neissner and S. -M. Niemi and J. W. Nightingale and C. Padilla and S. Paltani and F. Pasian and K. Pedersen and W. J. Percival and V. Pettorino and S. Pires and G. Polenta and M. Poncet and L. A. Popa and L. Pozzetti and F. Raison and R. Rebolo and A. Renzi and J. Rhodes and G. Riccio and E. Romelli and M. Roncarelli and E. Rossetti and B. Rusholme and R. Saglia and Z. Sakr and A. G. Sánchez and D. Sapone and B. Sartoris and M. Sauvage and J. A. Schewtschenko and M. Schirmer and P. Schneider and M. Scodeggio and A. Secroun and G. Seidel and M. Seiffert and S. Serrano and P. Simon and C. Sirignano and G. Sirri and J. Skottfelt and L. Stanco and J. Steinwagner and P. Tallada-Crespí and D. Tavagnacco and A. N. Taylor and H. I. Teplitz and I. Tereno and S. Toft and R. Toledo-Moreo and F. Torradeflot and I. Tutusaus and L. Valenziano and J. Valiviita and T. Vassallo and G. Verdoes Kleijn and A. Veropalumbo and Y. Wang and J. Weller and A. Zacchei and G. Zamorani and F. M. Zerbi and E. Zucca and M. Bolzonella and C. Burigana and R. Cabanac and L. Gabarra and M. Huertas-Company and V. Scottez and D. Scott},
      year={2025},
      eprint={2503.15335},
      archivePrefix={arXiv},
      primaryClass={astro-ph.GA},
      url={https://arxiv.org/abs/2503.15335}, 
}

@ARTICLE{Choi_2016,
       author = {{Choi}, Jieun and {Dotter}, Aaron and {Conroy}, Charlie and {Cantiello}, Matteo and {Paxton}, Bill and {Johnson}, Benjamin D.},
        title = "{Mesa Isochrones and Stellar Tracks (MIST). I. Solar-scaled Models}",
      journal = {\apj},
         year = 2016,
        month = jun,
       volume = {823},
       number = {2},
          eid = {102},
        pages = {102},
          doi = {10.3847/0004-637X/823/2/102},
archivePrefix = {arXiv},
       eprint = {1604.08592},
 primaryClass = {astro-ph.SR},
       adsurl = {https://ui.adsabs.harvard.edu/abs/2016ApJ...823..102C}
}

@ARTICLE{Paczynski_1986,
       author = {{Paczynski}, B.},
        title = "{Gravitational Microlensing by the Galactic Halo}",
      journal = {\apj},
         year = 1986,
        month = may,
       volume = {304},
        pages = {1},
          doi = {10.1086/164140},
       adsurl = {https://ui.adsabs.harvard.edu/abs/1986ApJ...304....1P}
}

@ARTICLE{van_Dokkum_2018,
       author = {{van Dokkum}, Pieter and {Danieli}, Shany and {Cohen}, Yotam and {Merritt}, Allison and {Romanowsky}, Aaron J. and {Abraham}, Roberto and {Brodie}, Jean and {Conroy}, Charlie and {Lokhorst}, Deborah and {Mowla}, Lamiya and {O'Sullivan}, Ewan and {Zhang}, Jielai},
        title = "{A galaxy lacking dark matter}",
      journal = {\nat},
         year = 2018,
        month = mar,
       volume = {555},
       number = {7698},
        pages = {629-632},
          doi = {10.1038/nature25767},
archivePrefix = {arXiv},
       eprint = {1803.10237},
 primaryClass = {astro-ph.GA},
       adsurl = {https://ui.adsabs.harvard.edu/abs/2018Natur.555..629V}
}

@article{Trujillo_2019,
    author = {Trujillo, Ignacio and Beasley, Michael A and Borlaff, Alejandro and Carrasco, Eleazar R and Di?Cintio, Arianna and Filho, Mercedes and Monelli, Matteo and Montes, Mireia and Rom獺n, Javier and Ruiz-Lara, Tom獺s and S獺nchez?Almeida, Jorge and Valls-Gabaud, David and Vazdekis, Alexandre},
    title = {A distance of 13 Mpc resolves the claimed anomalies of the galaxy lacking dark matter},
    journal = {Monthly Notices of the Royal Astronomical Society},
    volume = {486},
    number = {1},
    pages = {1192-1219},
    year = {2019},
    month = {03},
    issn = {0035-8711},
    doi = {10.1093/mnras/stz771},
    url = {https://doi.org/10.1093/mnras/stz771},
    eprint = {https://academic.oup.com/mnras/article-pdf/486/1/1192/28390837/stz771.pdf},
}

@misc{nikjoo2026weaklensingphotometricdensity,
      title={Weak Lensing by Photometric Density Ridges}, 
      author={Mehraveh Nikjoo and Joe Zuntz and Ben Moews},
      year={2026},
      eprint={2603.04025},
      archivePrefix={arXiv},
      primaryClass={astro-ph.CO},
      url={https://arxiv.org/abs/2603.04025}, 
}

@article{Baltz_2004,
   title={Microlensing Candidates in M87 and the Virgo Cluster with the<i>Hubble Space Telescope</i>},
   volume={610},
   ISSN={1538-4357},
   url={http://dx.doi.org/10.1086/382071},
   DOI={10.1086/382071},
   number={2},
   journal={The Astrophysical Journal},
   publisher={American Astronomical Society},
   author={Baltz, Edward A. and Lauer, Tod R. and Zurek, David R. and Gondolo, Paolo and Shara, Michael M. and Silk, Joseph and Zepf, Stephen E.},
   year={2004},
   month=Aug, pages={691–706} }

@misc{tang2026newmeasurementsdistancesgalaxies,
      title={New Measurements of Distances to Galaxies in the NGC 1052 Field with the Hubble and James Webb Space Telescopes: Testing the Bullet-Dwarf Origin of the Trail}, 
      author={Yimeng Tang and Gagandeep S. Anand and Aaron J. Romanowsky and Pieter G. van Dokkum and Kevin A. Bundy},
      year={2026},
      eprint={2606.05144},
      archivePrefix={arXiv},
      primaryClass={astro-ph.GA},
      url={https://arxiv.org/abs/2606.05144}, 
}

@ARTICLE{Williams_2024,
       author = {{Williams}, Liliya L.~R. and {Kelly}, Patrick L. and {Treu}, Tommaso and {Amruth}, Alfred and {Diego}, Jose M. and {Li}, Sung Kei and {Meena}, Ashish K. and {Zitrin}, Adi and {Broadhurst}, Thomas J. and {Filippenko}, Alexei V.},
        title = "{Flashlights: Properties of Highly Magnified Images Near Cluster Critical Curves in the Presence of Dark Matter Subhalos}",
      journal = {\apj},
         year = 2024,
        month = feb,
       volume = {961},
       number = {2},
          eid = {200},
        pages = {200},
          doi = {10.3847/1538-4357/ad1660},
archivePrefix = {arXiv},
       eprint = {2304.06064},
 primaryClass = {astro-ph.CO},
       adsurl = {https://ui.adsabs.harvard.edu/abs/2024ApJ...961..200W}
}

@article{Wright_2000,
doi = {10.1086/308744},
url = {https://doi.org/10.1086/308744},
year = {2000},
month = {may},
publisher = {},
volume = {534},
number = {1},
pages = {34},
author = {Wright, Candace Oaxaca and Brainerd, Tereasa G.},
title = {Gravitational Lensing by NFW Halos},
journal = {The Astrophysical Journal}
}

@ARTICLE{NFW,
       author = {{Navarro}, Julio F. and {Frenk}, Carlos S. and {White}, Simon D.~M.},
        title = "{The Structure of Cold Dark Matter Halos}",
      journal = {\apj},
         year = 1996,
        month = may,
       volume = {462},
        pages = {563},
          doi = {10.1086/177173},
archivePrefix = {arXiv},
       eprint = {astro-ph/9508025},
 primaryClass = {astro-ph},
       adsurl = {https://ui.adsabs.harvard.edu/abs/1996ApJ...462..563N}
}







   
  




\end{document}